\documentclass[sigconf]{acmart}

\AtBeginDocument{%
  }

\copyrightyear{2025} 
\acmYear{2025} 
\setcopyright{acmlicensed}\acmConference[CHI '25]{CHI Conference on Human Factors in Computing Systems}{April 26-May 1, 2025}{Yokohama, Japan}
\acmBooktitle{CHI Conference on Human Factors in Computing Systems (CHI '25), April 26-May 1, 2025, Yokohama, Japan}
\acmDOI{10.1145/3706598.3713185}
\acmISBN{979-8-4007-1394-1/25/04}

\usepackage{subcaption}
\usepackage{stfloats}

\begin{document}

\title
[Blending The Worlds: World-Fixed Visual Appearances in Automotive AR] 
{Blending the Worlds: World-Fixed Visual Appearances in Automotive Augmented Reality} 

\author{Robin Connor Schramm}
\orcid{0000-0002-4775-4219}
\affiliation{
  \institution{Mercedes-Benz Tech Motion GmbH}
  \city{B{\"o}blingen}
  \country{Germany}
}
\affiliation{
  \institution{RheinMain University of Applied Sciences}
  \city{Wiesbaden}
  \country{Germany}
}
\email{robin.schramm@mercedes-benz.com}

\author{Markus Sasalovici}
\orcid{0000-0001-9883-2398}
\affiliation{
  \institution{Mercedes-Benz Tech Motion GmbH}
  \city{B{\"o}blingen}
  \country{Germany}
}
\affiliation{
  \institution{Ulm University}
  \city{Ulm}
  \country{Germany}
}
\email{markus.sasalovici@mercedes-benz.com}

\author{Jann Philipp Freiwald}
\orcid{0000-0002-1977-5186}
\affiliation{
  \institution{Mercedes-Benz Tech Motion GmbH}
  \city{B{\"o}blingen}
  \country{Germany}
}
\email{jann_philipp.freiwald@mercedes-benz.com}

\author{Michael Otto}
\orcid{0000-0003-0212-0965}
\affiliation{
  \institution{Mercedes-Benz AG}
  \city{Stuttgart}
  \country{Germany}
}
\email{michael.m.otto@mercedes-benz.com}

\author{Melissa Reinelt}
\orcid{0000-0002-3110-3673}
\affiliation{
  \institution{Mercedes-Benz AG}
  \city{Stuttgart}
  \country{Germany}
}
\email{melissa.reinelt@mercedes-benz.com}

\author{Ulrich Schwanecke}
\orcid{0000-0002-0093-3922}
\affiliation{
  \institution{RheinMain University of Applied Sciences}
  \city{Wiesbaden}
  \country{Germany}
}
\email{ulrich.schwanecke@hs-rm.de}
\renewcommand{\shortauthors}{Schramm et al.}

\begin{abstract}
    With the transition to fully autonomous vehicles, non-driving related tasks (NDRTs) become increasingly important, allowing passengers to use their driving time more efficiently. In-car Augmented Reality (AR) gives the possibility to engage in NDRTs while also allowing passengers to engage with their surroundings, for example, by displaying world-fixed points of interest (POIs). This can lead to new discoveries, provide information about the environment, and improve locational awareness. To explore the optimal visualization of POIs using in-car AR, we conducted a field study (N = 38) examining six parameters: positioning, scaling, rotation, render distance, information density, and appearance. We also asked for intention of use, preferred seat positions and preferred automation level for the AR function in a post-study questionnaire. Our findings reveal user preferences and general acceptance of the AR functionality. Based on these results, we derived UX-guidelines for the visual appearance and behavior of location-based POIs in in-car AR.
\end{abstract}

  \ccsdesc[500]{Human-centered computing~Empirical studies in HCI}
  \ccsdesc[500]{Human-centered computing~Mixed / augmented reality}
  \ccsdesc[500]{Human-centered computing~HCI design and evaluation methods}
  \ccsdesc[500]{Human-centered computing~Field studies}

\keywords{Augmented Reality, Point of Interest, POI, In-Car, Visualization, Vehicle, Passenger, Automotive User Interfaces}

\begin{teaserfigure}
  \includegraphics[width=\textwidth]{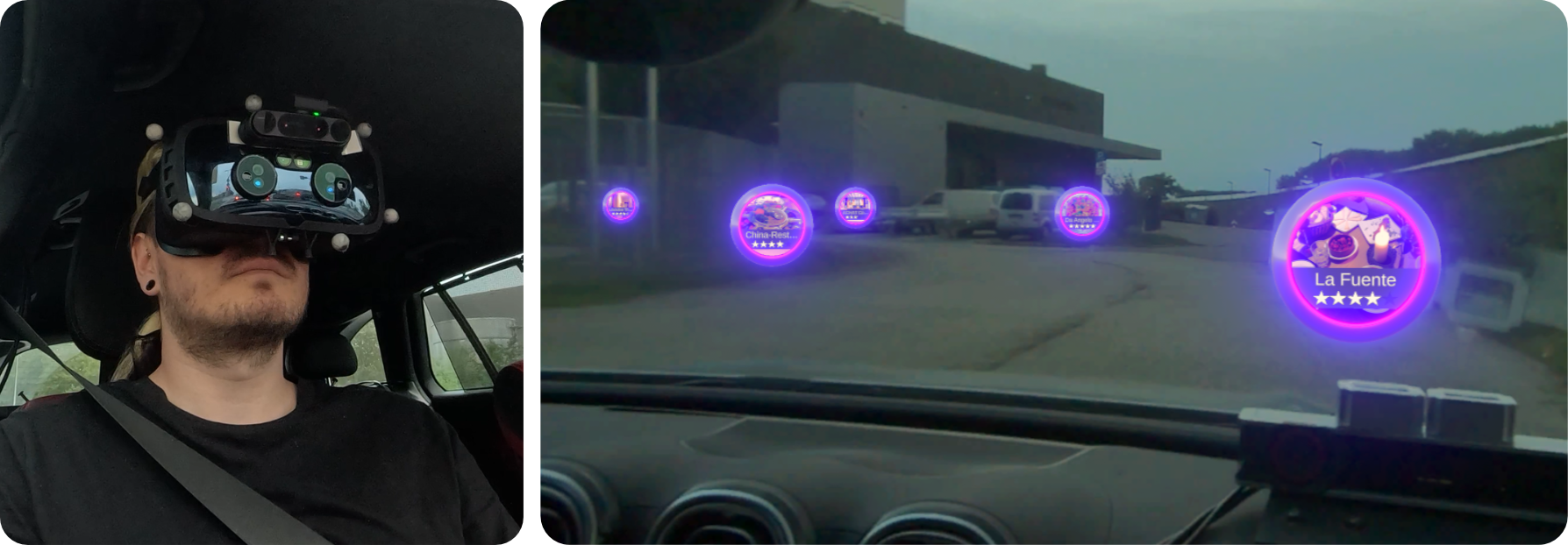}
  \caption{Blending the Worlds enables passengers to explore their surroundings through digital points of interest displayed in Augmented Reality. We use the Varjo XR-3 with additional optical tracking (left). Points of interest are visualized as spheres outside the vehicle (right).}
  \Description{The setup for our in-car augmented reality system is shown. In the left image, a person is seated in a car, wearing the Varjo XR-3 headset. In the right image, the view inside the Varjo XR-3 is displayed, showing the scene through the car's windshield. Outside the car are five spheres representing points of interest, with restaurant names and images on them.}
  \label{fig:teaser}
\end{teaserfigure}

\maketitle

\section{Introduction}
\label{section:introduction}
The integration of Augmented Reality (AR) in the automotive sector is rapidly expanding. Major car manufacturers are incorporating AR features like Head-Up Displays and video-based AR to provide drivers and passengers with navigation guidance and contextual information \cite{Elhattab23AutomotiveAR}. This expansion offers opportunities for enhanced navigation, information delivery, and entertainment. Simultaneously, advancements in head-worn AR-hardware, including improvements in field of view (FoV) and display quality, are making Head-Mounted Displays (HMDs) and AR glasses increasingly feasible for in-car use \cite{riegler2021augmented}. Such technologies could increase the passenger experience by providing access to information about the outside environment \cite{BergerGridStudyInCarPassenger2021}. As such, modern cars offer a space for a wide array of non-driving-related activities such as communication, information access, and digital media consumption \cite{MatsumuraActivePassengering18}. Additionally, the rise of autonomous vehicles means that passengers will have more free time during transit \cite{mcgill2020challenges} e.g. for work \cite{Mathis2021work, medeiros2022shielding} or for leisure activities like gaming \cite{Togwell2022gaming}. 

With the importance of getting information about a passengers environment \cite{BergerGridStudyInCarPassenger2021, Mehrabian74EnvironmentalPsychology, Pfleging16NDRNeeds, BergerRearSeatDoor21}, future in-car AR systems could incorporate points of interest (POIs) to offer real-time, context-specific information about the car's surroundings overlaid onto the real-world. POIs provide users digital representations of real places such as restaurants, parks, or gas stations \cite{Psyllidis2022POIs} and can play important roles in supporting various human activities \cite{sun2023conflating}. Specific to the automotive context, POIs could also include traffic conditions, hazards, landmarks, and locations of interest. This integration of digital content with real-world context could also facilitate easier understanding of information at a glance \cite{haeuslschmid2016design} or could sharpen the riding experience \cite{MatsumuraActivePassengering18}.

To explore the optimal visualization of POIs using in-car AR we developed \textit{Blending The Worlds}, a system to display location-based data in the environment of a real moving vehicle via a pass-through HMD. The purpose of this system is to give users a way to explore their environment while using in-car AR as a passenger. We investigate possible ways to position, scale, and visualize POIs for this use-case. Based on our system, we want to answer the following research questions:

\begin{itemize}
    \item \textbf{RQ1:} \textit{What appearance and positioning of POIs do users prefer in an in-car AR context?}
    \item \textbf{RQ2:} \textit{Do users accept an AR system inside a moving vehicle?}
\end{itemize}

We answer these questions through a field study (N=38). Participants assessed multiple parameters for the conditions height, size, rotation, render distance, information density, and appearance. In addition, we gathered quantitative and qualitative data regarding the user acceptance and the user intention for the in-car AR-function. Based on our results, we provide insights into the design principles and best practices for optimizing the appearance and positioning of POIs in AR environments. Our system is designed primarily for passenger use, with a focus on rear-seat passengers, an underexplored area in automotive user interfaces \cite{BergerRearSeatDoor21}. However, our findings may also apply to other passenger positions. The main contributions of this paper are as follows:
\begin{itemize}
    \item A comprehensive examination of measures for POI positioning, rotation, scaling, render distance, information density, and appearance within automotive AR environments through a field study with 38 participants.
    \item Investigation of user acceptance for a HMD-based in-car AR system for displaying POIs.
    \item Guidance for deriving UX guidelines for visual appearances and behavior of location-based POIs for in-car AR.
\end{itemize}
\section{Related work}
\label{section:related}
Our work explores the visualization and placement of world-fixed data for in-car AR in form of POIs. As such, our related work consists of the topics of augmented passenger experiences and visualization of location-based data in AR.

\subsection{Augmented passenger experiences}
In 2021, Berger et al. \cite{BergerGridStudyInCarPassenger2021} explored major factors for a convenient in-car passenger experience. They identified that the outside environment had a major impact on passengers' perceived convenience. They brought this in line with past investigations on showing information about surrounding attractions \cite{BergerTactile19, MatsumuraActivePassengering18}. They also highlighted the enormous potential of innovative technologies like AR to be used by passengers. In addition, they also explored the importance of information access for convenient user experiences. They stated that users want to receive information beyond traditional ones like the time of arrival or the speed level. Future in-car applications should incorporate information about the direct surroundings, preferably based on passenger preferences and needs.

Similarly, Matsumura and Kirk \cite{MatsumuraActivePassengering18} explored the use of interactive car window systems to enhance passenger engagement with the external environment during car journeys. Their study identified several key themes that contribute to an improved passenger experience, including active participation, reflective interaction, social connectivity, and temporal awareness. User could could freeze the view to the outside and clip a frame using a gesture. As such, users could essentially capture or create a POI to seek further information or to review their journey later.

Challenges can arise in the utilization of HMDs for passengers in vehicles, which have to be overcome. McGill et al. \cite{mcgill2020challenges} identified crash safety, motion sickness (MS), social acceptance, and interaction in constrained spaces as challenges to overcome. Nevertheless they advocated for further exploration of HMDs in vehicle environments since they can potentially improve passenger experience in terms of productivity, entertainment and isolation. Similarly, Riegler et al. \cite{riegler2020agenda} proposed to investigate the use of HMDs further, especially in the context of autonomous driving. For example, they advocated to combat MS and create practical in-car work- and entertainment related experiences. 

In addition, in-vehicle human-computer interaction (HCI) studies are typically evaluated using simulators like virtual reality setups \cite{ColleySwiVR22}, CAVE systems \cite{SawitzkyCAVE23}, or professional driving simulators. While virtual driving simulators have their merits, they can't fully replace real-world testing \cite{petterson2019virtually}. For instance, participants might behave differently in a simulator, which for example lacks the stress of actual traffic conditions \cite{gomaa2020studying}.

Regarding specifically AR-based POI exploration in cars, Berger et al. \cite{BergerRearSeatDoor21} presented a design concept of an interactive car door for displaying POIs for rear-seats. They implemented a digital prototype and conducted a user study remotely with a simulation on a computer display. Participants were shown attractions along a route with which they could interact using a touch panel to receive additional information. The functionality to get more details about POIs was highly appreciated by all participants and most of them could have imagined to use it frequently. Notably, participants chose the rear seat over the front seat for their interactive car door.  However, their design for the location-based POIs was limited to the attractions name and a simple line. As a future step they highlighted the necessity for testing the system in a real-world experiment.

In summary, previous research has highlighted the potential of AR for enhancing in-car passenger experiences. However, most prior studies have relied on simulations, digital prototypes, or did not incorporate head-worn AR. In contrast, our work advances the field by conducting a real-world experiment using a VST-HMD, demonstrating how AR can enhance passenger experiences by presenting location-based information.

\subsection{Visualization of location-based Data}
There is extensive related work exploring the visualization of spatial content in AR. In this section, we explore related work regarding AR-visualization techniques, depth perception for outdoor AR, and occlusion handling in outdoor AR. However, findings from studies focused on stationary or pedestrian contexts may have limited relevance to our application, as aligning the car's position with POI locations introduces positional errors.

Zollmann et al. \cite{Zollmann2021ArVisTechniques} provide a taxonomy for visualization techniques in AR. They define six dimensions for design spaces in AR visualization. These are purpose, visibility, virtual cues, filtering, abstraction, and compositing. For our work, the purposes dimension is the most relevant, as we explore the presentation of virtual content being integrated into the real world while being part of an in-car user interface. Zollmann et al. \cite{Zollmann2021ArVisTechniques} found, that each visualization technique in their related work addresses one or more of five aspects: "a) achieving visual coherence, b) a better depth perception, c) reducing information clutter, d) supporting exploration, and e) directing attention." \cite{Zollmann2021ArVisTechniques}. In Section \ref{sec:design}, we explain our design decisions based on these dimensions and aspects. 

One example of an investigation into AR visualization in outdoor environments is the work by Hertel and Steinicke \cite{SteinickeArMaritimePois2021}. They conducted an experiment on egocentric depth perception in large-distance outdoor settings (up to 75 meters) using optical see-through (OST) AR. Their findings indicate that rendering drop shadows below objects hovering over a surface and employing bright, warm colors enhances depth perception accuracy. They further noted that good visibility is crucial for effortless and precise depth perception in outdoor scenarios, where bright colors are most effective. 

Additionally, they identified a relationship between object size and error rates, with larger objects associated with reduced errors in depth perception. However, while this relationship was statistically significant, the effect size was minimal, making it negligible for practical applications. Although Hertel and Steinicke \cite{SteinickeArMaritimePois2021} aimed to apply their findings to maritime environments, their main study was conducted in an empty parking lot, rendering their results potentially relevant to our use case of in-car AR. However, their focus on OST AR, which has inherent limitations, may not directly translate to the video see-through (VST) AR systems employed in our study.

Another study examining AR in outdoor environments was conducted by Ghaemi et al. \cite{Ghaemi23ARPlacement}, focusing on the effect of visualization placement on user experience during outdoor augmented data tours. The study utilized situated data visualizations in AR and compared three placement strategies: directly on buildings, on the ground near the user, and on a virtual map displayed in front of the user. Visualizations placed directly on buildings were the most preferred by participants. However, the findings are limited to the specific context of outdoor environments and may not generalize to indoor or in-car settings. Furthermore, the study did not address the design or appearance of the visualized objects. Their findings may also not directly translate to our system as they used an OST AR device.

There are also works that explore depth perception for AR applications. Diaz et al. \cite{Diaz2017depthPerception} conducted two experiments in an indoor environment investigating the effect of shading, cast shadows, aerial perspective, texture, dimensionality, and billboarding on users depth perception of virtual objects. They found cast shadows to be highly beneficial for depth estimation. The use of billboarding and larger virtual objects also had significant impact on depth perception, however not as much as the use of shadows. 

Livingston et al. \cite{Livingston2009IndoorOutdoorAr} examined depth perception of virtual objects in indoor and outdoor environments using a depth matching task. The indoor environment featured strong linear perspective cues, which the researchers sought to replicate outdoors. Depth was generally underestimated indoors, consistent with patterns observed in virtual environments, while outdoors, participants tended to overestimate depth. The synthetic perspective cues used outdoors were moderately effective, leading to reduced depth estimates for distant objects.

As for occlusion in outdoor AR visualization, Kasapakis and Gavalas \cite{kasapakis2017occlusion} developed a system using geolocative raycasting to detect surrounding buildings in realtime. Thus, a realistic FoV for players of an AR game could be generated. Skinner et al. \cite{Skinner2018IndirectPoiBrowser} proposed an indirect AR browser to address availability challenges by combining crowdsourced, precaptured street-level imagery with geospatial data. Their work demonstrates how this browser annotates buildings and landmarks in the user's environment and examines its feasibility by analyzing the approach's performance.

In summary, extensive research has explored spatial content visualization in AR, primarily in stationary or pedestrian contexts. However, these findings may have limited applicability to in-car AR, where aligning a moving vehicle's position with POI locations introduces positional errors. Prior work on outdoor AR visualization has addressed occlusion handling through techniques such as geolocative raycasting and indirect AR browsers that rely on pre-captured imagery and geospatial data. While effective in static settings, these approaches do not account for the unique challenges of dynamic, vehicle-based environments. In contrast, our work focuses on real-time AR visualization in moving vehicles, specifically addressing positional inaccuracies and continuously shifting perspectives when displaying location-based data.
\section{System}
\label{sec:system}
Blending the Worlds enables passengers of a moving vehicle to explore their surroundings through digital POIs displayed in AR. Here, POIs are visualized as spheres outside the vehicle, as shown in Figure \ref{fig:teaser}. A video of the system is available in the supplementary material.

Our POIs fit the \textit{Label} pattern described by Lee et al. \cite{Lee24SituatedVisAR} for categorization of situated visualizations in AR. Labels are designed to provide additional context to referents, offering observers insights into aspects of the physical environment that are not easily accessible through conventional means. They also suggest that labels have the potential to become a key feature driving the success of AR in the near future. The widespread adoption of AR labels could have an impact on daily life comparable to the influence of spontaneous Wikipedia searches on everyday conversations. AR labels can also be dynamically optimized regarding placement and appearance. As such, we made our POIs customizable regarding the parameters height, size, rotation, render distance, and information density.

\begin{figure}[h]
    \centering
    \includegraphics[width=\linewidth]{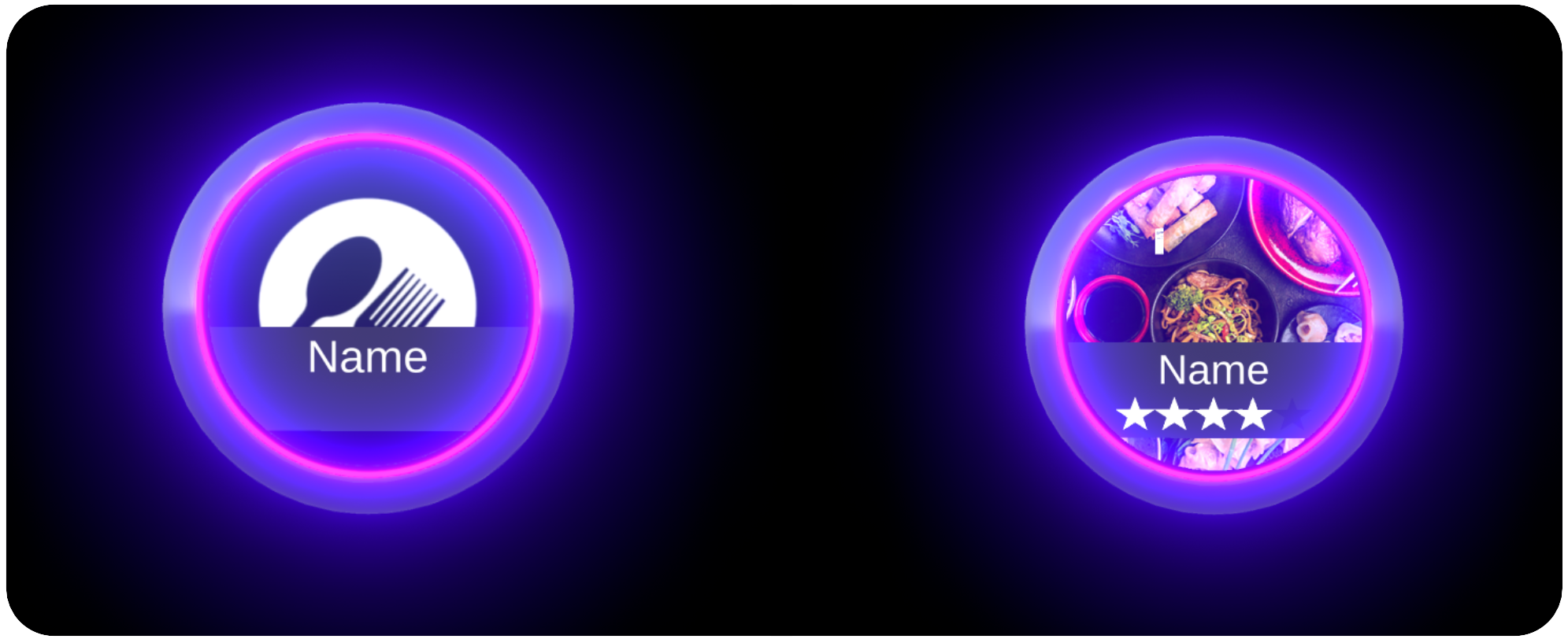}
    \Description{The image displays two spheres set against a black background. Each sphere features a grey reflective outer border, with a smaller inner border that resembles a glowing neon tube, emitting a bright pink light. A bloom effect radiates from the glowing border of both spheres. Each sphere depicts a point of interest (POIs). The left sphere contains an icon of a spoon and fork, symbolizing a restaurant. The right sphere showcases an image of various types of Asian food. Both spheres also include a dark grey bar across the bottom third, displaying the name of the respective POI. Additionally, the right sphere features a four-star rating beneath the name.}
    \caption{Our proposed POI visualization in front of a black background. The left POI shows an icon representing a restaurant. The right POI shows a sample restaurant image and a rating from zero to five stars.}
    \label{fig:POI_Appearances}
\end{figure}

\subsection{Design}
\label{sec:design}
The design of our POIs is shown in Figure \ref{fig:POI_Appearances}. The core design features a sphere with a grey reflective outer border, with a smaller inner border that resembles a glowing neon tube, emitting a bright pink light. This visual design adheres to three key purposes defined by Zollmann et al. \cite{Zollmann2021ArVisTechniques}: visual coherence, exploration, and directing attention. Visual coherence is achieved by aligning the design with the aesthetic of our test vehicle, which incorporates similar styles in its interior and user interface. Exploration is achieved by providing contextual information for exploring the scene through the  information displayed on the POIs. Additionally, the design fulfills the purpose of directing attention, standing out in most environments due to its distinct, unnatural appearance, color, and form. A bright color is also recommended by Hertel and Steinicke \cite{SteinickeArMaritimePois2021}, especially for large distances in outdoor AR.

The other two key purposes defined by Zollmann et al. \cite{Zollmann2021ArVisTechniques} are clutter reduction and depth perception. Regarding clutter reduction, each of the five adjustable parameters for POIs (height, size, rotation, render distance, and information density) has the potential to influence clutter, as detailed in the following sections. Displaying more content provides additional information for the user; however, excessive content can lead to clutter, which may increase cognitive load \cite{kim2011multidimensional}. To support depth perception, our system primarily employs changing object size as a depth cue \cite{goldstein2009sensation}. Additionally, in some conditions, we utilize changing object height to indicate depth as well \cite{goldstein2009sensation}. Although drop shadows are frequently used to improve depth estimation for AR objects, most studies on AR visualization have been conducted in controlled indoor settings or open outdoor spaces \cite{erickson2020reviewOSTAr, Zollmann2021ArVisTechniques}. However, in our system, POIs are often rendered above or on top of buildings rather than on flat surfaces like streets. In such scenarios, the use of shadows could disrupt visual coherence. Moreover, pose estimation for a moving vehicle is significantly more complex than for a stationary observer or someone walking \cite{McGill22PassengXR}. This could lead to wrong positioning of shadows, potentially hindering depth estimation.

\subsection{Height}
\label{sec:system_height}
POIs could be placed floating above their respective locations to indicate popular tourist destinations \cite{Lee24SituatedVisAR}, directly on buildings on street level, or directly on the street in front of buildings \cite{Ghaemi23ARPlacement}. POIs placed on street level may help to better estimate their position and respective buildings. In contrast, virtual POIs floating above their locations may help reduce clutter and acts as a depth cue while still conveying the existence of an interestsing location.

We define two adjustable parameters for POIs' vertical position: base height and dynamic height scaling. This allows for POIs to be displayed at any desired height and for optional distance-based height scaling. The base height determines the initial elevation of POIs above the ground. For POIs with \textit{static} height, no additional vertical scaling is applied. However, for \textit{dynamic} POIs, the height increases based on the distance to the user. This way, closer POIs are still placed on their target locations, while far POIs float above their target locations. Figure \ref{fig:HeightSizeDescription} shows a simplified illustration of the POIs' height behavior. The dynamic scaling is realized by using the POIs base height, the \textit{POI distance} and three additional parameters: a \textit{minimum distance threshold}, a \textit{maximum (max.) distance threshold}, and a \textit{maximum (max.) scaling}. The vertical position of POIs beyond the \textit{minimum distance threshold} is increased beyond their base height via the calculated \textit{scale} parameter from Equation \ref{eq:scalingFormula}.
\begin{equation} 
    \label{eq:scalingFormula}
        \text{scaling} = \left(\frac{\text{POI distance}}{\text{max. distance threshold}}\right)^2\cdot\text{max. scaling}
\end{equation}

\subsection{Size}
\label{sec:system_size}
AR labels can be dynamically optimized in their appearance for fitting the observer's information needs, e.g. by adjusting their scale \cite{Lee24SituatedVisAR}. Larger POIs direct more attention, while smaller POIs could reduce clutter. We define two adjustable parameters for POI size, similar to height: base size and dynamic scaling. The base size determines the POI spheres radius in meters. For statically sized POIs, the base size remaines unaltered. Consequently, POIs with static sizes appear smaller depending on their distance from the user, analogous to real-world objects. Dynamic scaling of size was achieved similarly to the dynamic scaling of vertical position, where we applied the scale calculated from Equation \ref{eq:scalingFormula} to the base size. Dynamically sized POIs still appear smaller the further away they are from the user, just with a lesser effect. Figure \ref{fig:HeightSizeDescription} shows a simplified illustration of the POIs' size behavior.

\subsection{Rotation}
\label{sec:system_rotation}
Our POIs consist of 3D models, consequently they can be rotated and face the user in different ways. With a \textit{billboarding} behavior, POIs rotate around the x- and y-axes (using a left handed coordinate system) to continuously face the user. Alternatively, our POIs can maintain their original orientation without any rotational adjustments. This entails no rotation around the x-axis and a y-rotation aligning the POIs face almost parallel to the street, akin to a street sign. Consequently, with no rotation, observers can see the sides and empty backs of POIs. This can potentially help to judge the side of the street a POI is on. In addition, this directs attention to POIs on the user's current street, as only their faces with further information is visible. Simultaneously, this could help reduce clutter, as POIs pointing in different directions still show an interesting location while not overwhelming the user with the bright circle, images, and text. Figure \ref{fig:RotationRenderdistanceDescription} shows the two rotation behaviors.

\subsection{Render Distance}
\label{sec:system_renderdistance}
Clutter from overlapping POIs needs to be taken into account, e.g. by reducing the amount of objects shown at once \cite{Lee24SituatedVisAR}. Thus, POIs in our system can be disabled at a certain distance with a minimum and maximum threshold. After crossing a threshold, POIs begin to fade in or out. The distance over which POIs fade can also be defined for both nearby and far POIs. The fading mitigates POIs abruptly appearing at the far edge of the render distance. Figure \ref{fig:RotationRenderdistanceDescription} shows an example for two render distances.

\subsection{Information Density}
\label{sec:system_informationdensity}
Labels are not limited to textual content and can contain independent visualizations or other content \cite{Lee24SituatedVisAR}. We display the name, a star-rating and an image on our POIs. In addition, the color of the glowing border is also changeable. Each type of content can also be disabled depending on the use-case. Adjusting the content of POIs could potentially influence visual clutter. Figure \ref{fig:POI_Appearances} shows two possible configurations for POI content.

\subsection{Hardware}
We use the Varjo XR-3 pass-through HMD (shown in Figure \ref{fig:teaser}) due to its specifications\footnote{Varjo Technologies Oy: Varjo XR-3, the first true mixed reality headset. \url{https://varjo.com/products/varjo-xr-3/} (accessed on 12.08.2024)} and compatibility with middleware from LP-Research\footnote{LPVR Middleware a Full Solution for AR / VR. \url{https://www.lp-research.com/middleware-full-solution-ar-vr/} (accessed on 12.09.2024)}. The Varjo XR-3 features a resolution of 70 pixels per degree in its focus area \cite{Kappler22VarjoEvaluation}, a FoV of 115, a refresh rate of 90Hz, and a pass-through latency of less than 20ms. It also supports six degrees of freedom (6-DoF) HMD tracking in a moving vehicle using middleware from LP-Research supported by an additional car-mounted inertial measurement unit. The Varjo XR-3 was connected to a desktop computer (CPU: Intel® Core™ i7-12700K,
RAM: Kingston Fury Beast 32 GB 3200 Mhz DDR4, Mainboard: Asus Z690 TUF gaming, Graphics Card: Gainward GeForce RTX 4080 Phoenix). The computer was secured in the trunk of the vehicle.
\section{User Study}
\label{section:study}
In this Section, we outline the pre-study and main user study aimed to assess parameters for visualizing POIs on an AR device based on our system described in Section \ref{sec:system}. This investigation is particularly focused on the unique context of a moving vehicle, an area that has not been extensively explored yet. The visualization parameters we examined encompassed height, size, rotation, render distance, information density, and appearance. Additionally, we explored the acceptance and intention of using AR technology in moving vehicles, specifically for the purpose of displaying POIs. We formed several hypotheses for each parameter described in Section \ref{sec:independentVariables}. 

We conducted an exploratory pilot study to establish default values for each independent variable. This pilot study employed the same apparatus as described in Section \ref{sec:apparatus} for the main study. For participants, we recruited five individuals with expertise in HCI and immersive technologies, selected through convenience sampling.

The procedure closely followed the approach detailed in Section \ref{sec:procedure} for the main study, with the primary modification being the omission of all questionnaires. Instead, participants were given the ability to adjust the parameters that constituted the variables outlined in Section \ref{sec:independentVariables} using interactive sliders. Each variable was adjusted individually and sequentially, following the order specified in Section \ref{sec:independentVariables}. The vehicle continued driving on our study track until the participant was satisfied with the adjustments for all sliders. The adjustable parameters for each condition were as follows:
\begin{itemize}
    \item \textbf{Height:} Base height, minimum distance threshold, maximum distance threshold, and maximum height scaling.
    \item \textbf{Size:} Base size, minimum distance threshold, maximum distance threshold, and maximum size scaling.
    \item \textbf{Rotation:} No sliders, just a choice between billboarding and no rotation.
    \item \textbf{Render distance:} Far edge distance, fading distance, far threshold, and near threshold.
    \item \textbf{Information Density:} Participants could turn on or off the name and the star rating. They also could toggle between an image and an icon.
\end{itemize}

For the analysis, we calculated the average values between the five participants, which correspond to \textit{low\_static} height, \textit{small\_static} size, \textit{billboarding} rotation, \textit{long distance} for render distance, and \textit{high information density}. The POIs in Figure \ref{fig:teaser} represent these default values.

\subsection{Participants}
\label{sec:participants}
A total of 38 participants were recruited for the main-study, consisting of 20 males and 18 females, with an average age of 40.9 years (\textit{range: 20 to 61 years}). Among the participants, eleven ($28.95\%$) had no prior experience with AR, having never used AR glasses or HMDs. Ten participants ($26.32\%$) reported minimal experience, having engaged with AR apps or games on their smartphones. Fourteen participants ($37\%$) had limited exposure to AR glasses, using them 1-3 times, while 3 participants ($7.89\%$) were classified as experienced users, regularly utilizing such devices. Prior to the study, participants were asked to wear contact lenses if they required prescription eyewear. This recommendation aimed to ensure consistent comfort and to eliminate confounding the factor of hardware limitations as much as possible, as most glasses don't fit inside the Varjo XR-3 HMD. Among the participants, 27 individuals ($ 71.05\%$) did not require any prescription, while the remaining 11 participants ($28.95\%$) adhered to the recommendation and used contact lenses during the study. As such, our study encompasses a diverse range of participants regarding age and familiarity with AR systems.

Participants were also asked to indicate how frequently they experience MS while engaging in secondary tasks as a passenger in a moving vehicle. They could answer on a 5-point likert scale ranging from \textit{never} to \textit{(almost) always}. Fourteen participants ($36.84\%$) reported never experiencing MS, nine participants ($23.68\%$) reported rare occurrences, eleven ($28.95\%$) reported occasional sickness, and four ($10.53\%$) reported experiencing MS often or almost always. None of the participants aborted a study session due to MS.

\subsection{Apparatus}
\label{sec:apparatus}
Participants were positioned in the right rear seat of a midsize sedan. The front right seat was adjusted to its forwardmost position to allow for optimal head-tracking and to ensure the participants' safety. Participants used an Xbox Elite Wireless Controller as the input device for responding to questionnaires. Two additional occupants accompanied the participants during the study. Apart from the driver, the experiment conductor occupied the left rear seat of the vehicle. Positioned there, the experiment conductor could view the participant's perspective on a screen and documented all relevant observations throughout the study. To ensure realistic and controlled driving conditions, we chose a private, industrial area with moderate traffic, including other vehicles and pedestrians. To maintain uniform driving conditions, the car's speed limiter was set to the maximum allowable speed within the study environment, capped at 30 km/h.

\begin{figure}[ht]
    \centering
    \includegraphics[width=.6\linewidth]{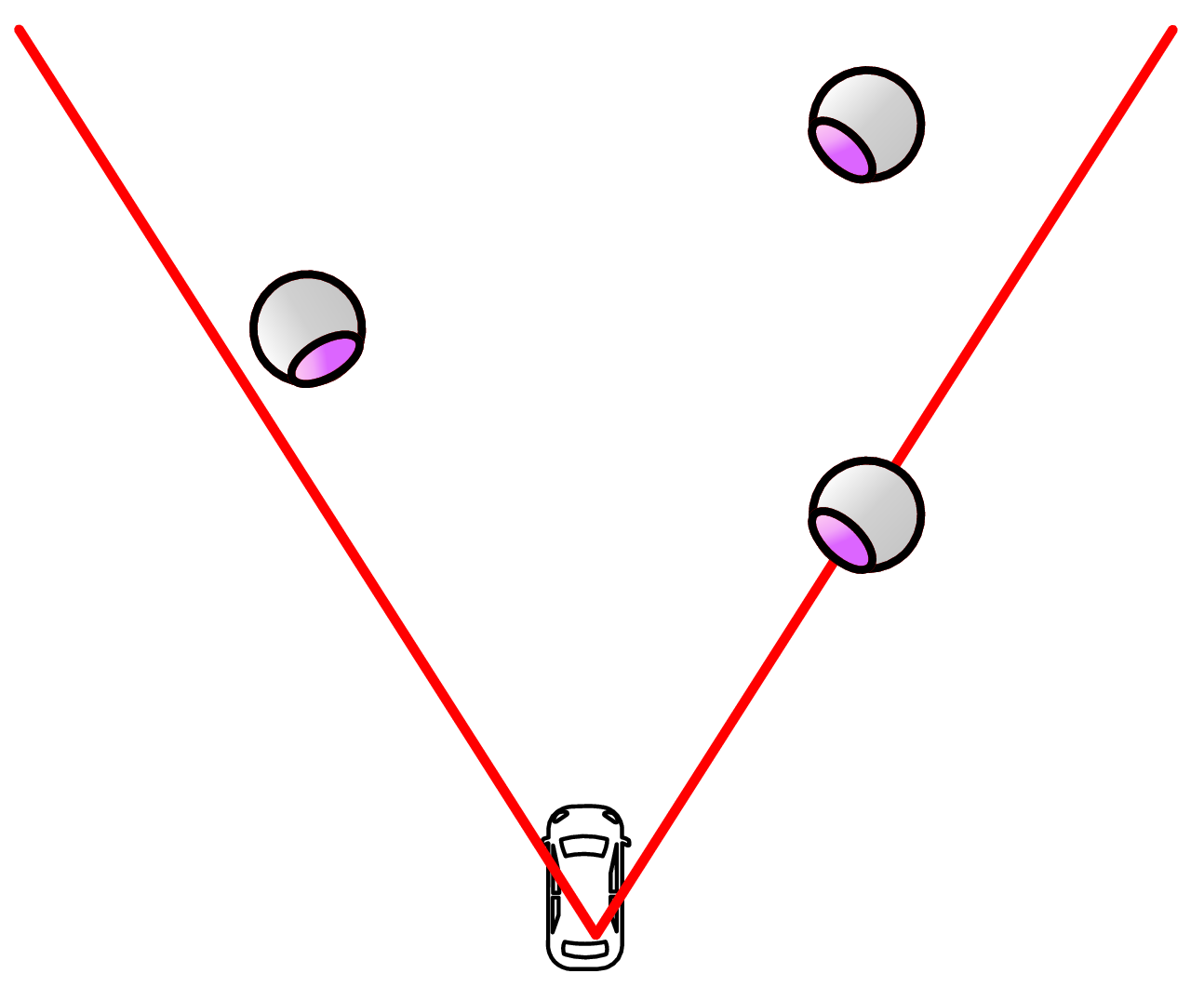}
    \Description{A schematic illustration of the participants' field of view during the study. Two red lines originating from a small cars' are shown. The lines have a 115 degree angle between them. In front of the car, three POIs are shown, one on the left, two on the right. The right on near the car is only partially inside the red lines field of view. The other two POIs are inside the field of view.}
    \caption{Schematic visualization of participants' field of view during the study. The red lines show the 115\textdegree{} field of view of the Varjo XR-3. The relative position and scale of the POIs and the car in the image are true to scale.}
    \label{fig:TopdownSchematics}
\end{figure}

\subsection{Procedure}
\label{sec:procedure}
A complete study session for one participant took approximately one and a half hours. The questionnaires are located in their entirety in the appendix. At the beginning of the study session, the participant was welcomed and taken to the designated study environment. First, the participant provided informed consent regarding the management of their privacy and personal data. Subsequently, the study conductor delivered a presentation on the basics of AR and POIs utilizing presentation slides. This was followed by the pre-questionnaires, which are described in further detail in Section \ref{sec:measures}. Afterwards, the participant could enter the study vehicle and was driven to the study's starting point. There, they were briefed on the procedural aspects of the in-car study and instructed on the operation of the HMD. Additionally they were informed about the potential for MS, including the procedures to follow in the event of experiencing such symptoms. Afterwards, they put on the HMD and were given the controller to start a round of acclimatization with the AR function activated. During the acclimatization, the HMDs pass-through mode was activated and POIs were displayed next to the street while the car was driving one lap through the study environment. For the acclimatization, the default values described in Section \ref{sec:independentVariables} and shown in Figure \ref{fig:teaser} were used. Next, the study conductor read out the user story to the participant. The scenario depicted a potential future situation in which the user, accompanied by two colleagues, is on a business trip, driving through a city with a significant distance remaining in their journey. In their quest to find a nearby place for lunch, the user utilizes a new AR function, requesting the vehicle to display restaurants along their route. Subsequently, they can observe relevant targets within their environment.

After this, the study procedure began. During each round, POIs were shown outside the car alternating between the left and right side of the road. The POIs were placed in world-space, with each of them possessing specific lat-long coordinates derived from the street's position. The POIs resembled restaurants as described in the user story with varying information, positioning, and appearance depending on the study state. The restaurants shown on the POIs consisted of simulated data and did not correlate with the buildings seen in the real environment, since the study took place in a private industrial area with no real restaurants nearby. Instead, the POIs were equally distributed with alternating positions to the left and right sides of the street. POIs were placed with a distance between five and ten meters measured from the center of the street. Figure \ref{fig:TopdownSchematics} shows the relative position and scale of the POIs and the car together with the field of view the participants had without moving their head.

\subsection{Independent Variables}
\label{sec:independentVariables}
There were varying predefined states for each of our independent variables: height, size, rotation, render distance, and information density. For the study's default values, the mean values of the pilot study described in Section \ref{section:study} were used. 
For each round, participants were told which independent variables were modified and on what they needed to concentrate on. However, they were not told in what way the variables were modified. The order for the independent variable categories was consistent for each participant and followed the order of the following paragraphs. The conditions within the categories were counterbalanced using latin square.

\begin{figure*}
    \centering
    \begin{subfigure}[b]{.49\textwidth}
        \centering
        \includegraphics[width=\textwidth]{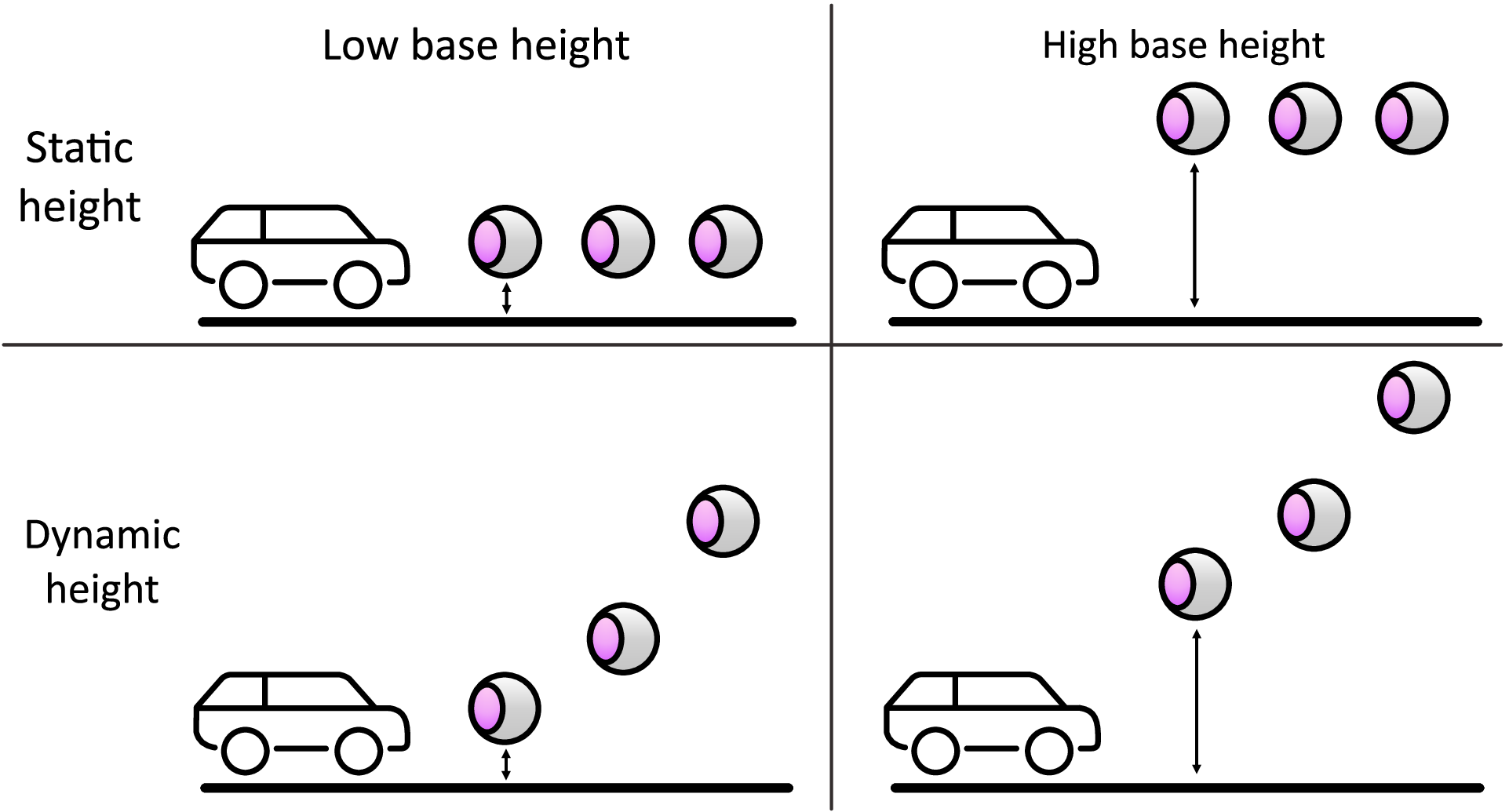}
    \end{subfigure}
    \hfill
    \begin{subfigure}{.488\textwidth}
        \centering
        \includegraphics[width=\textwidth]{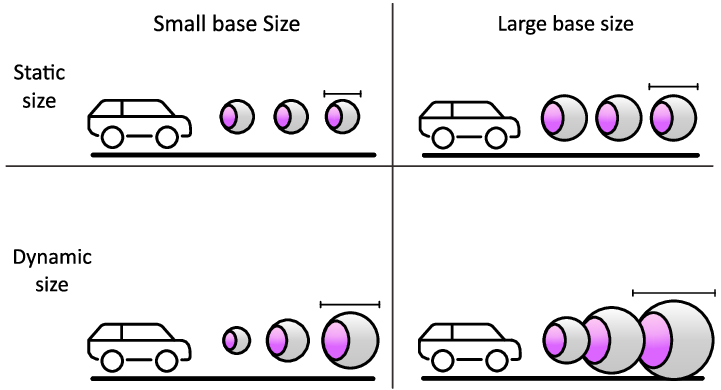}
    \end{subfigure}
    \caption{Schematic representation of POI placement and appearance in the four study conditions regarding height (left) and size (right).}
    \Description{A schematic illustration of the height and size conditions. The Figure consists of two images, each showing a two by two matrix. Each of the eight squares in the two matrices shows a 2D graphic of a car in front of three points of interest each. On the left matrix, the POI height is displayed. An arrow in each image visualizes how the points of interest differ in their vertical position. Low base height coupled with static height shows the points of interest at the height of the car. High base height coupled with static height shows the points of interest above the car. Low base height coupled with dynamic height shows the points of interest on a curved line, starting the car's height and going above the car. High base height coupled with dynamic height shows the points of interest on a curved line, starting above the car and getting even higher the further away they are from the car. On the right, the POI size is displayed. A scale in each image visualizes how the points of interest differ in their size. Small base size coupled with static size shows the points of interest, roughly sized as two thirds of the car. Large base size coupled with static size shows the points of interest roughly as large as the car. Small base size coupled with dynamic size shows the points of interest getting bigger the further away they are from the car. They start a little smaller than the small base size and get larger than the large base size. Large base size coupled with dynamic size shows the points of interest again scaling depending on the distance to the car. They start as large as the car and get almost doubled in size.}
    \label{fig:HeightSizeDescription}
\end{figure*}

\subsubsection*{\textbf{Height}}
The POIs' height attribute is described in Section \ref{sec:system_height}. For the study, we adjusted both the base height and the dynamic scaling. Figure \ref{fig:HeightSizeDescription} illustrates the four conditions for height \textit{low\_static}, \textit{low\_dynamic}, \textit{high\_static}, and \textit{high\_dynamic}. Our dependent variables for height are satisfaction, visibility, and pleasantness. The low base height conditions positioned POIs at the user's eye-level while the high base height conditions set POIs to hover 15 meters above ground-level. For all dynamic height trials, the minimum distance threshold equaled to 30 meters, the maximum distance threshold to 500 meters, and the maximum height scaling to 100 meters. Those values were based on the satisfaction factor in our pilot study. For POI distances smaller than the minimum distance threshold the \textit{scaling} equaled 1, for POI distances larger than the maximum distance threshold the \textit{scaling} equaled the maximum height scaling. 

We expected the conditions with low base height to be the preferred conditions, as the deployed Varjo XR-3 weights around 980g\footnote{\label{foot:Varjo}\url{https://varjo.com/products/varjo-xr-3/} (accessed on 12.08.2024)} and can potentially cause head strain while being used in a moving car \cite{Schramm23Assessing}. As such, the placement of POIs at approximately eye-level could be more comfortable. Also, with a low base height, there should be a more direct association with digital POIs and the real world. As such, we hypothesized that POIs with a low base level lead to higher visibility. Dynamic scaling of POIs should help orient the user and help decluttering the FoV, potentially leading to higher pleasantness. However, this could come at the cost of a lower association between POIs and their location in the real world. In contrast, a high base height could reduce visual clutter on eye-level while giving a broad overview over the nearby POIs. Hence, we assumed the following hypotheses for \textit{height}:
\begin{itemize}
    \item $H_{H1}$: A base height of approximately eye level leads to higher satisfaction.
    \item $H_{H2}$: A base height of approximately eye level leads to higher visibility.
    \item $H_{H3}$: Dynamic height scaling leads to higher pleasantness.
\end{itemize}

\subsubsection*{\textbf{Size}}
The POIs' size attribute is described in Section \ref{sec:system_size}. For the study, we adjusted the base size and the dynamic scaling. Figure \ref{fig:HeightSizeDescription} illustrates the our four size conditions \textit{small\_static}, \textit{small\_dynamic}, \textit{large\_static}, and \textit{large\_dynamic}. Our dependent variables for size are satisfaction, visibility, and pleasantness. A low base size equated to a POI diameter of 3 meters, while a large base size equated to 7.5 meters. For all dynamic size trials, the minimum distance threshold equaled to 50 meters, the maximum distance threshold to 500 meters, and the maximum size scaling factor to 7. The values are based on the satisfaction factor in our pilot study.

We expected the conditions with low base size to be the preferred conditions with the highest satisfaction and pleasantness, as they don't obstruct a big portion of the outside view and thus could positively impact the experience \cite{BergerGridStudyInCarPassenger2021}. Additionally we hypothesized that the larger base size leads to higher visibility, as the larger POIs could improve readability, especially for POIs that are located further away. Also, the dynamic scaling could improve satisfaction and visibility for POIs across all distances as they adapt based on the distance to the user. They still appear smaller the further away they are, showing somewhat realistic behavior while not overly cluttering the FoV. Thus, we assumed the following hypotheses:
\begin{itemize}
    \item $H_{S1}$: A POI base size of approximately three meters leads to higher \textit{satisfaction} and \textit{pleasantness}.
    \item $H_{S2}$: POIs with a large base size of approximately 7.5 meters lead to higher \textit{visibility}.
    \item $H_{S3}$: Dynamic size scaling leads to higher \textit{visibility} and \textit{pleasantness}.
\end{itemize}

\begin{figure*}
    \centering
    \begin{subfigure}[b]{.55\textwidth}
        \centering
        \includegraphics[width=\textwidth]{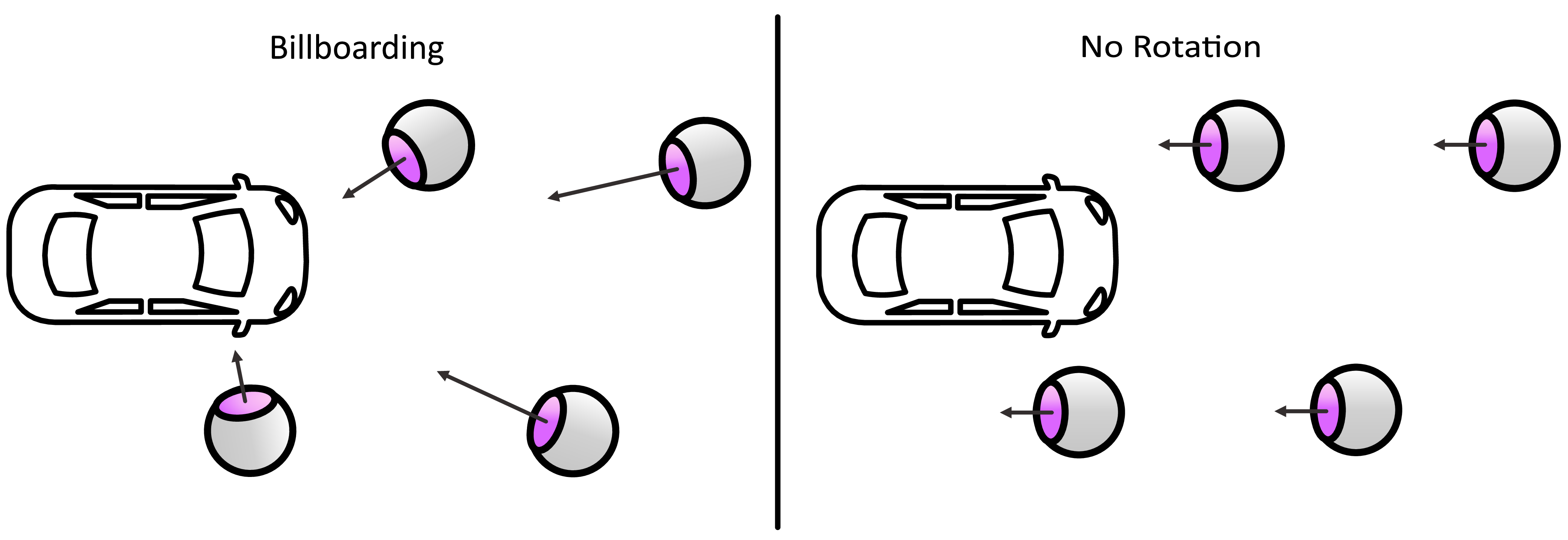}
        \Description{}
    \end{subfigure}
    \hfill
    \begin{subfigure}{.4\textwidth}
        \centering
        \includegraphics[width=\textwidth]{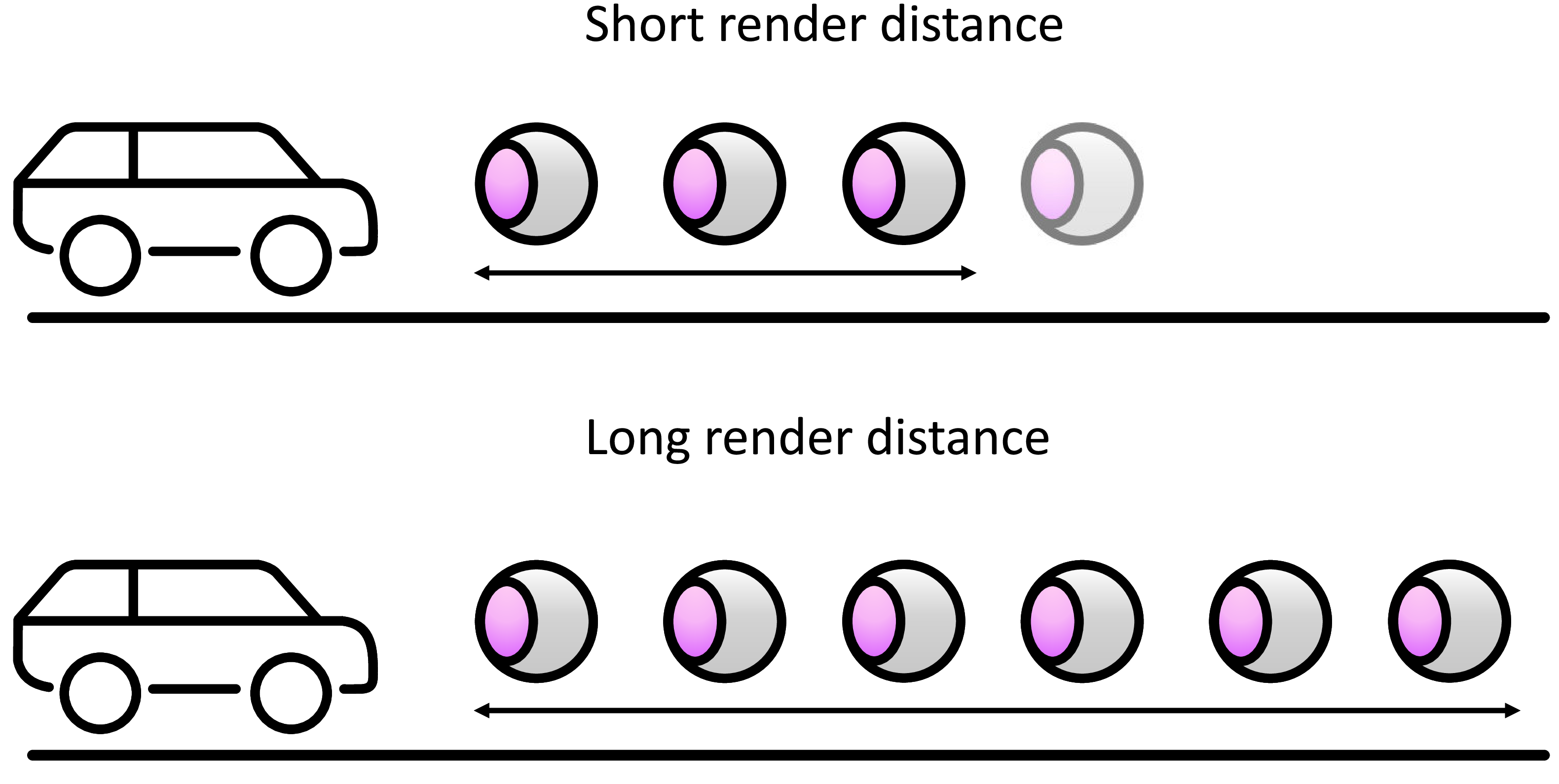}
        \Description{A schematic illustration of the rotation and render distance conditions. The Figure consists of four images. Two images on the left represent the rotation condition. Each shows a car with four points of interest from above. For billboarding, all four points of interest are rotated towards the car's passenger seat, indicated by four arrows. For no rotation, each of the four points of interest a rotated paralell to the car's direction, again indicated by four arrows. The two images on the right represent the render distance. Each of them shows a car with points of interest from the side. For the short render distance, four points of interest are shown, where the last one is half transparent. For the long render distance, six points of interest are shown.}
    \end{subfigure}
    \caption{Schematic representation of the two study conditions manipulating the POIs' rotation (left) and render distance (right).}
    \label{fig:RotationRenderdistanceDescription}
\end{figure*}

\subsubsection*{\textbf{Rotation}}
The POIs' rotation is described in Section \ref{sec:system_rotation}. We tested two conditions regarding rotation in the study: \textit{billboarding} and \textit{no rotation}, as illustrated in Figure \ref{fig:RotationRenderdistanceDescription}. For the rating of the rotation, we used four word pairs for clarity, support, complexity, and pleasantness. The word pairs were taken from the short version of the User Experience Questionnaire \cite{schrepp2017design} with the exception of pleasantness, which we formulated ourselves.
  
We expected the billdboarding behavior to be the more supportive condition since there the POI-content is consistently available and readable. This could improve information delivery and thus be perceived as more pleasant to use. The non-rotating POIs may be easier to understand, as they resemble the static behavior known from real-life street signs. Additionally, they could support users by conveying information about the streets' direction and provide a clearer association between POIs and streets. As such, our hypotheses are as follows:
\begin{itemize}
    \item $H_{R1}$: Billdboarding POIs are more supportive for delivering information and are thus more pleasant.
    \item $H_{R2}$: Non-rotating POIs are easier to understand and give a clearer overview of the environment.
\end{itemize}

\subsubsection*{\textbf{Render Distance}}
The render distance described in Section \ref{sec:system_renderdistance} had two conditions in the study: \textit{long distance} and \textit{short distance}, as illustrated in Figure \ref{fig:RotationRenderdistanceDescription}. For the \textit{long distance} condition, the far edge to fade-in POIs was chosen at a distance of 500 meters. This resulted in all existing POIs to be displayed at all times, since our study environment had a maximum lenght of around 450 meters. For the \textit{short distance} condition, the far egde was set to 150 meters, resulting in three to four POIs being visible simultaneously while traversing a straight street segment. The fading distance was set to 50 meters for the far threshold and to 2.5 meters for the near threshold. For the render distance, participants could rate the POIs' time of appearance, ranging from \textit{way too early} to \textit{way too late}.

For render distance, we expected the \textit{short distance} to be preferable due to the reduced visual clutter and the difficulity to read far away POIs. Thus, our hypothesis regarding this variable is:
\begin{itemize}
    \item $H_{RD1}$: A short render distance is preferred by users.
\end{itemize}

\subsubsection*{\textbf{Information Density}}
There were two levels of content, resulting in two conditions tested: \textit{Low information density} and \textit{high information density}. Here, our dependent variable was satisfaction. For the \textit{low information density} condition, only a generic restaurant icon and the restaurant name was displayed. For the \textit{high information density} condition POIs showed the restaurant's name, an image, and a star rating ranging from zero to five. Examples for both conditions can be seen in Figure \ref{fig:POI_Appearances}. There was also the possibility to rate each part of the POIs' content individually during the post-questionnaires. Participants were shown a 3x4 matrix with the x-axis being the POI components name, star rating, icon, and image. The y-axis comprised of points in time on when the components could be shown: always, when nearby, and never. The matrix, including the results, is illustrated in table \ref{tab:InformationMatrix}.

We expected the \textit{high information density} to be the preferred condition, since it provides the most relevant information at a glance without requiring any additional interaction. Additionally, we hypothesized that more information should be displayed for near POIs, since far away POIs are less readable.

\begin{itemize}
    \item $H_{I1}$: POIs with three types of data are preferred by users.
    \item $H_{I2}$: Users want to have more data displayed for nearby POIs.
\end{itemize}

\subsubsection*{\textbf{Appearance}}
Our POIs comprise of spheres with one cut side, allowing for a flat space to display 2D information on. While facing the user in the billboarding conditions, the POIs appeared as two dimensional objects, as seen in Figure \ref{fig:POI_Appearances}. Most of the POIs face is filled with either a representative image stemming from the real location or a icon indicating the POIs' category. We chose a deliberately artifical look for the POIs' appearance, form, and color to make them stand out against the environment and to be visually interesting. As such, our hypotheses regarding appearance are as follows:
\begin{itemize}
    \item $H_{A1}$: The visual presentation of our POIs is appealing.
    \item $H_{A2}$: The POIs' color makes them stand out against the environment. 
\end{itemize}

\subsection{Measures}
\label{sec:measures}
The participants were administered questionnaires before, during, and after the in-car study. The questionnaires that were specifically formulated for our study will be described on an abstract level in this Section and are located in their entirety in the appendix. All relevant results for the questionnaires are reported in Section \ref{section:results}, with detailed results also found in the appendix.

\subsubsection*{\textbf{Pre-Questionnaires}}
The pre-questionnaires included a socio-demographic questionnaire and two questionnaires regarding the participants affinity for both new technologies in general, as well as their affinity for AR. The pre-questionnaires were administered in a separate room and were filled out via keyboard and mouse on a computer. 
The socio-demographic questionnaire included questions about the participants' gender, age, body height, AR experience, their need for prescription glasses, and their suspectibility for MS. The results of the socio-demographic questionnaire are reported in Section \ref{sec:participants}.
For technical affinity, we used the affinity technology interaction scale (ATI) from Franke et al. \cite{franke2019personal}. For questions regarding participants' affinity specifically for AR, we used a mixture of self formulated questions and questions from Janzik \cite{janzik2022studie} with slight modifications.

\subsubsection*{\textbf{Study-Questionnaires}}
During the in-car part of the user study, participants could fill out the questionnaires for the dependent variables through an AR-interface included in the study software without the need of removing the HMD. The questions were always filled out while the car was in standstill, since the UI for the questions occluded most of the HMD's FoV and thus could induce MS \cite{Sasalovici23ArMs}. The questions were tailored to the conditions and mostly used seven-point Likert scales. In addition, all relevant verbal comments that the participants made during the in-car part of the study were noted in connection to the currently viewed condition. Those comments were later condensed into comment categories, which then got sorted into positive, neutral, and negative categories. The ratings for each condition are described respectively in Section \ref{sec:independentVariables}.

\subsubsection*{\textbf{Post-Questionnaires}}
After the study procedure in the car, there was no post-interview, participants directly filled out the post-questionnaires. These consisted of the acceptance regarding the AR-function, the POI appearance, and the intention of use. Like the pre-questionnaires, the post-questionnaires were administered in a separate room and were filled out via keyboard and mouse on a computer. All the post-questionnaire questions were self formulated and are listed in the appendix.

First, participants could rate their acceptance of the used AR function in general, using seven point Likert scales. For instance, some questions revolved around determining whether the AR function is deemed useful, innovative, or exciting. Afterward, participants were provided with two free-text fields where they could express their opinions on what they particularly enjoyed and disliked about the AR functionality.

The next set of questions regarded the POIs appearance, form, and color with seven point Likert scales. If the participant gave a negative rating with three and below, they were asked to fill out a freetext field why they were dissatisfied with this aspect of the POIs appearance. Then, for all participants, there were two additional freetext fields. There the participants were asked if they want to change any aspects of the POIs appearance other than those listed before. The other field had space to add any additional information or data to the POIs content that was not in the study.

The last set of questions regarded the intention of use for the AR-function. The first question regarded the preferred seat position depending on the cars' automation level. As such, the possible answers consisted of a 3x4 matrix where one axis was the seat position (driver, codriver, backseatpassenger, none) and the other axis was the automation level (level 3, level 4 and above, no automation). Multiple answers were possible. Participants were shortly briefed on what the automation levels meant beforehand.
Then, participants could rate if they want to use the AR function for different kinds of POIs, like theatres or gas stations from a list of nine possibilities. Afterwards they could pose their own kinds of POIs. 
\section{Results}
\label{section:results}
We used a 2x2 repeated measures ANOVA for the interpretation of the height and size questionnaire results. \textit{Post hoc} tests were then conducted using Tukey's range test. For rotation, render distance, and information density, we used paired t-tests. The conditions decribed in chaper \ref{sec:independentVariables} make up the independent variables, and the questionnaire scores make up the dependent variables for our analysis. We assumed a 5\% significance level for both ANOVA and t-tests. We only report significant results for the questionnaires. The qualitative data is reported in Section \ref{section:discussion}. More comprehensive results can be found in the appendix.

\begin{figure*}[ht]
    \centering
    \includegraphics[width=\linewidth]{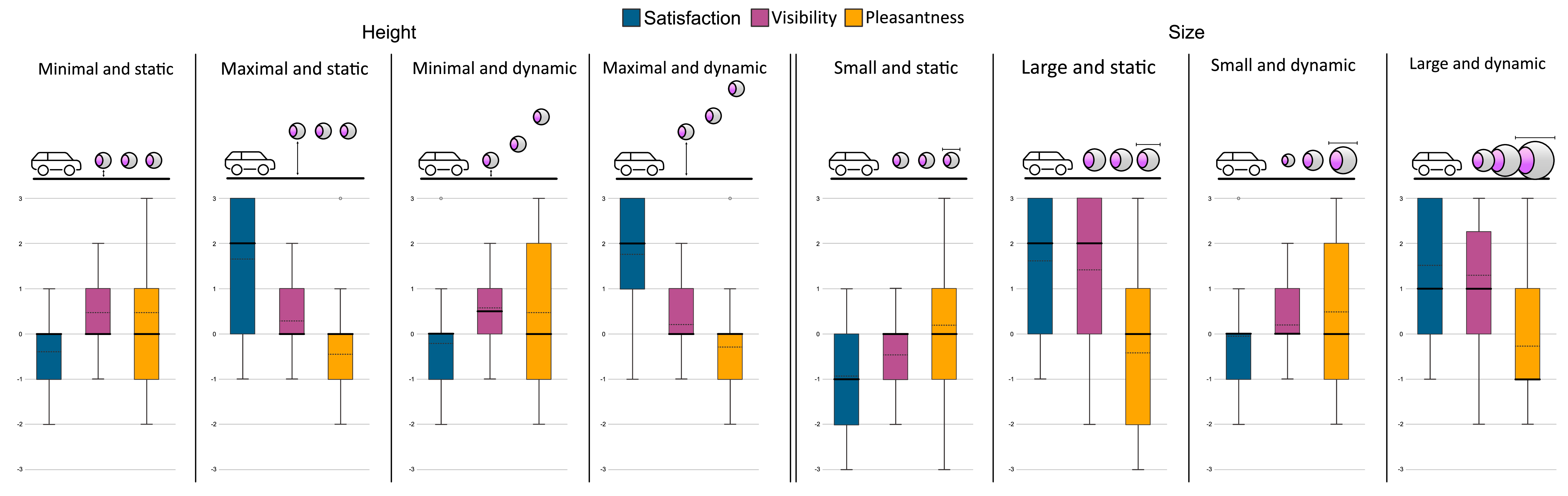}
    \Description{Eight Boxplots for ratings of height and size, including the dependent variables satisfaction, visibility, and pleasantness. On the left, four boxplots show the values for the four height conditions. The values for height are provided in Table 4 in the appendix. On the right, four boxplots show the values for the size conditions. The values for size are provided in Table 11 in the appendix.}
    \caption{Boxplots for scores regarding the height (left) and size (right) conditions. The bold lines indicate the median and the dotted lines indicate the mean. The dependent variables height/size, visibility, and pleasantness are shown grouped by the independent variables as illustrated above the graphs. For height, size, and visibility, values closer to zero are better. For pleasantness, higher values are better.}
    \label{fig:HeightSizeBoxplots}
\end{figure*}

\subsection{Height}
\label{sec:heightResult}
For the conditions manipulating the vertical position of POIs, we used the rating of satisfaction, visibility, and pleasantness as dependent variables, illustrated in Figure \ref{fig:HeightSizeBoxplots}.

The base height of POIs had a significant main effect on participants rating of \textbf{satisfaction} ($F(1,37)=98.68, p<.001$). The low base height at eye-level worked best here, both when paired with dynamicity  ($M = -0.21; SD = 1.143$), and without dynamicity ($M = -0.39; SD = 1.001$). \textit{Post hoc} tests revealed significant interactions ($p<.001$) for each possible pairing between low base height and high base height. 

The base height also had a significant effect regarding \textbf{pleasantness} ($F(1, 37)=7.59, p = .009$). Here, \textit{post hoc} tests only revealed a significant interaction ($p = 0.036$) between the conditions \textit{low\_static} ($M=0.32; Mdn=0.0; SD=1.44$) and \textit{high\_static} ($M=-0.45; Mdn=0.0; SD=1.37$).

\subsection{Size}
\label{sec:sizeResults}
For the conditions manipulating the size of POIs, we used the rating of satisfaction, visibility, and pleasantness as dependent variables, illustrated in Figure \ref{fig:HeightSizeBoxplots}.

The four conditions for size had a significant interaction effect of participants rating of \textbf{satisfaction} ($F(1, 37)=19.50, p<.001$). \textit{Post hoc} tests revealed significant interactions between each possible pairing ($p<.001$), except for the pair \textit{large\_static} - \textit{large\_dynamic} ($p=0.932$). Notably, the \textit{small\_dynamic} condition was rated closest to exactly right ($M=-0.05; Mdn=0; SD=1.012$). In addition, the conditions with larger baser-sizes were rated as overly overt.

Similarly for \textbf{visibility} there was a significant interaction effect ($F(1, 37)=8.10; p=0.007$). \textit{Post hoc} tests revealed significant interactions between each possible pairing, except for the pair \textit{large\_static} - \textit{large\_dynamic}. For visibility, the \textit{small\_dynamic} condition was also rated close to exactly right ($M=0.21; Mdn=0; SD=0.991$). 

The base size of POIs had a significant main effect on participants rating of \textbf{pleasantness} ($F(1, 37)=4.90, p=.03$). \textit{Post hoc} tests showed a significant difference between the \textit{small\_dynamic} and \textit{large\_dynamic} conditions. The \textit{small\_dynamic} condition was rated best with ($M=0.53; Mdn=0.5; SD=1.447$).

\subsection{Rotation}
\begin{figure}[ht]
    \centering
    \includegraphics[width=\linewidth]{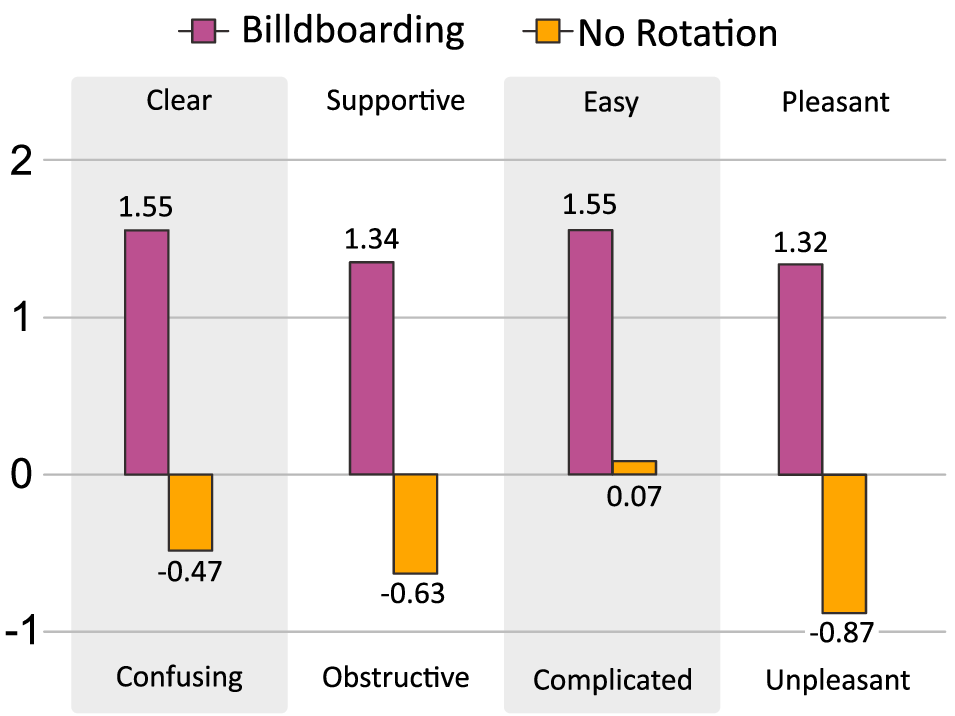}
    \Description{A word pair visualization for the rotation ratings. The ratings for billboarding are between one and two for each pair. The ratings for no rotation are located between zero and minus one, with the exception of complexity, which is located at zero point zero seven. The values are provided in table 18 in the appendix.}
    \caption{Mean values for word pairs regarding the rotation conditions. The scale ranged from -3 to 3. Higher is better.}
    \label{fig:RotationPairs}
\end{figure}

For the rating of conditions manipulating the rotation of POIs, we used four word pairs regarding clarity, support, complexity, and pleasantness as the dependent variables. The means for each pair and both conditions are illustrated in Figure \ref{fig:RotationPairs}. The \textit{Billboarding} condition was rated significantly higher than the \textit{No Rotation} condition for each wordpair ($p<.001$). 

\subsection{Render Distance}
For the render distance condition we used the fade-in timing as the dependent variable. A paired t-test showed significant differences between the two independent variables ($t(37) = 7.29; p < .001$). However, both timings were suboptimal. The \textit{long distance} was rated to render POIs too early ($M=-0.789; Mdn=-0.5; SD=1.32$) while for the \textit{short distance}, POIs were rendered too late ($M=0.684; Mdn=1.0; SD=1.07$).

\subsection{Information Density}
As for the content displayed on the POIs, the conditions had a significant main effect on the ratings ($t(36)=-10.46; p<.001$). The condition with \textit{low information density} had way too little content for the participants ($M=-2.27; Mdn=-3; SD=0.99$) and scored the worst possible rating in the median. The condition with \textit{high information density} was closer to optimal, but still had not enough content displayed for some participants ($M=-0.378; Mdn=0; SD=0.861$). Table \ref{tab:InformationMatrix} shows, which content participants want to have displayed at which time. Furthermore, participants could comment during the post-questionnaires for additional types of information that they want POIs to show.

\begin{table}[t]
    \centering
    \caption{The percentages of participants that wanted specific POI content types displayed at which points in time. Multiple selections were possible.}
    \label{tab:InformationMatrix}
    \begin{tabular}{c|c|c|c|c}
    \toprule
                & Name & Star Rating & Icon & Image \\
    \hline
    always      & 63\% & 68\%        & 40\% & 60\%  \\
    nearby      & 34\% & 21\%        & 20\% & 34\%  \\
    never       & 3\%  & 11\%        & 40\% & 5\%   \\
    \bottomrule  
    \end{tabular}
\end{table}

\subsection{Appearance}
The POIs general appearance ($M=5.11;,Mdn=5.0, SD=1.23$) and form ($M=5.21, Mdn=6.0, SD=1.19$) were rated highly. However, the color was polarizing ($M=4.32; Mdn=4.0; SD=1.58$) with 37\% of participants rating the color three and lower, while 47\% rated the color a five and better.

\subsection{Acceptance}
The participants' ratings for their acceptance of the AR-function are illustrated in Figure \ref{fig:acceptance}. The illustration only mentions the question categories, the complete questionnaire is located in the appendix. In addition, participants could make free comments on what they liked or disliked about the AR-function. The coded comment categories are shown in Table \ref{tab:AcceptanceComments}.

\subsection{Intention of Use}
\begin{figure*}[t]
    \centering
    \begin{subfigure}[t]{.49\textwidth}
        \centering
        \includegraphics[width=\textwidth]{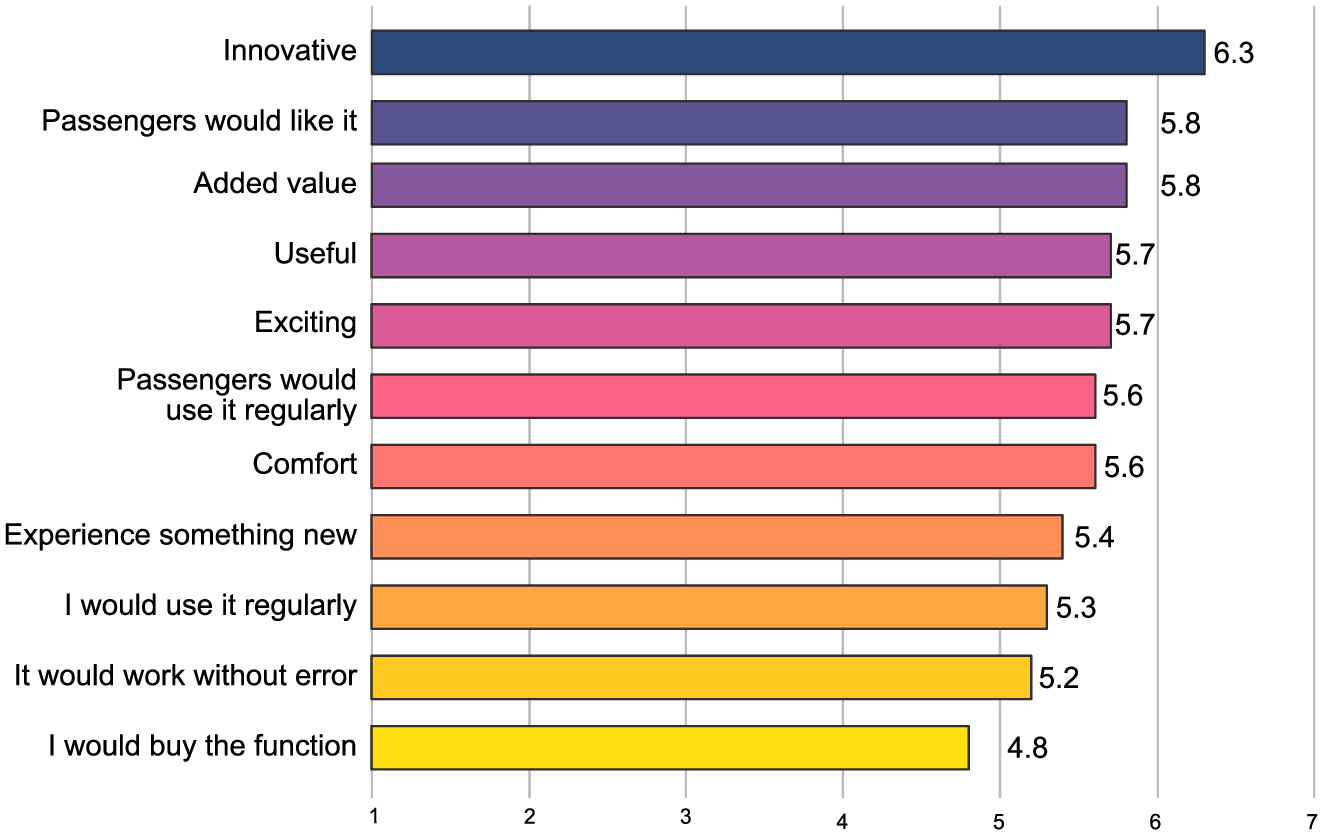}
        \caption{Acceptance}
        \label{fig:acceptance}
    \end{subfigure}
    \hfill
    \begin{subfigure}[t]{.49\textwidth}
        \centering
        \includegraphics[width=\textwidth]{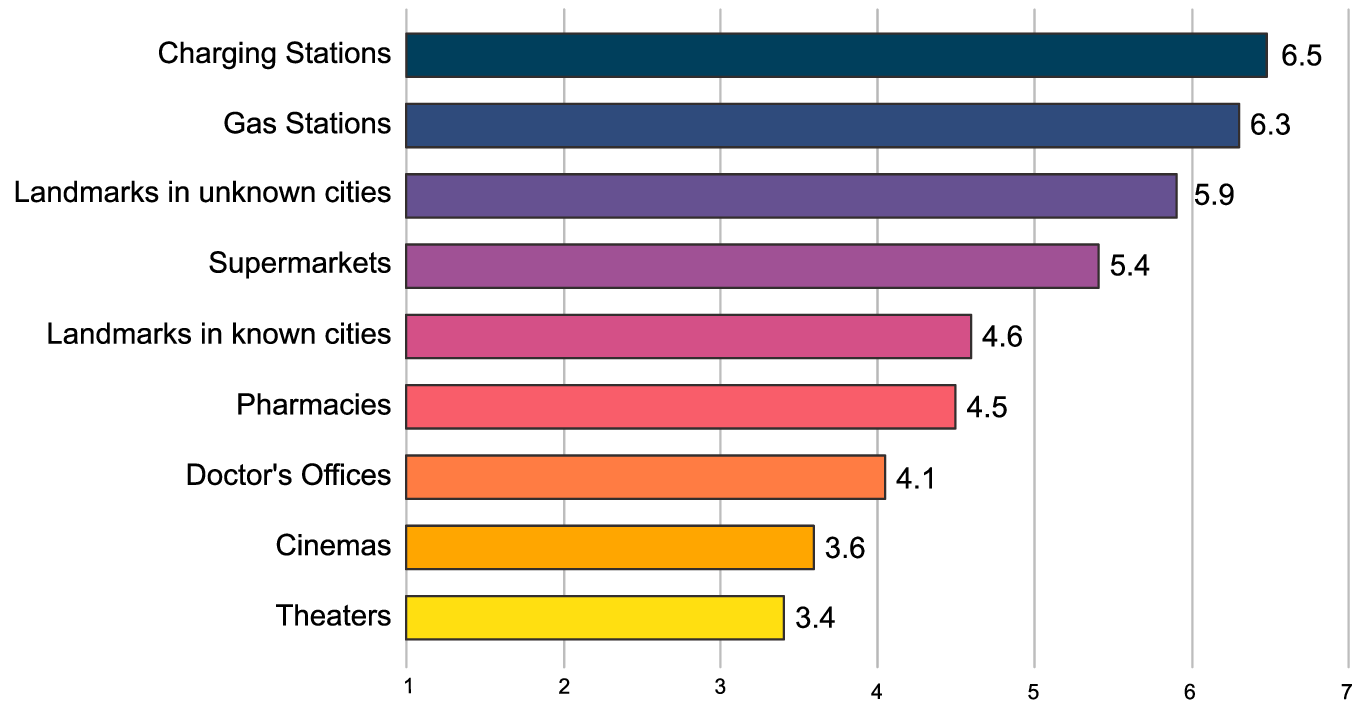}

        \caption{Intention of Use}
        \label{fig:intention}
    \end{subfigure}
    \caption{Participants' mean ratings regarding the acceptance (left) and their opinion on types of POIs that could be displayed of the AR-function (right). Higher is better.}
    \Description{Two barcharts. On the left, a bar chart showing the participants rating for acceptance of the AR-function. They are sorted by mean values. Innovative is at the top with 6.3, and "I would buy the function is at the bottom" with a mean of 4.8. The rest of ratings lies between 5.8 and 5.2. The values are provided in table 23 in the appendix. On the right, a bar chart showing the participants' ratings for the types of POIs that could be displayed, sorted by mean. It shows charging stations at 6.5, gas stations at 6.3, landmarks in unknown cities at 5.9, supermarkets at 5.4, landmarks in known cities at 4.6, pharmacies at 4.5, doctor's offices at 4.1, cinemas at 3.6, and theaters at 3.4.}
    \label{fig:AcceptanceIntention}
\end{figure*}

The last set of questions addressed the intention of use for the AR-function. Figure \ref{fig:intention} shows the participants's mean ratings for potential types of POIs. They could rate the question "Would you also like to use the POIs for the application scenarios mentioned below?" on a seven-point Likert scale. Participants could then mention own location types, which will be discussed in Section \ref{sec:intentionOfUse}. The matrix in Table \ref{tab:IntentionMatrix} illustrates the participants' preferred seat position depending on the cars' automation level for using the AR-function. The AR functionality is primarily being considered for passengers and other occupants besides the driver. The use of the AR-function while driving was mainly considered for higher automation levels. 

\begin{table}[t]
    \centering
    \caption{Percentage-matrix for the participants' preferred seat positions depending on the cars' automation level for using the AR-function. F-Passenger and B-Passenger stand for front-seat passenger and back-seat passenger respectively.}
    \label{tab:IntentionMatrix}
    \begin{tabular}{c|c|c|c|c}
    \toprule
                                   & Driver & F-Passenger & B-Passenger & Not at all \\
    \hline
    level 4                        & 87\%   & 92\%            & 87\%               & 3\%       \\
    level 3                        & 50\%   & 95\%            & 87\%               & 3\%       \\
    manual                         & 13\%   & 85\%            & 74\%               & 16\%      \\
    \bottomrule  
    \end{tabular}
\end{table}
\section{Discussion}
\label{section:discussion}
In this section, we discuss the results of our experiment in relation to the formulated hypotheses. The discussion will follow the order of tested parameters - height, size, rotation, render distance, information density and appearance. Furthermore, we summarize our discussed results in form of guidelines for designing POIs for in-car use, followed by user acceptance, user intention, and the limitations of our work.

\subsection{Height}
The first subsection of our experiment was focussed on the POIs' vertical positioning above ground level. Here, our results presented in Section \ref{sec:heightResult} confirm hypothesis $H_{H1}$ (A POI base height of approximately eye level leads to higher satisfaction). This result appears correlated with the perceived \textit{pleasantness} of the height setting, which likewise was rated most favorably for the base height setting on eye level. This is in line with best practices for spatial UI design\footnote{Riley Hunt. 2023. Spatial UI Design: Tips and Best Practices: \url{https://www.interaction-design.org/literature/article/spatial-ui-design-tips-and-best-practices} (accessed on 07.12.24)}, where the ideal content area starts slightly below eye-level. This claim is further supported by user comments, as the most counted commented categories for conditions with a low base height included "pleasant and fitting" ($N=19$) and "this height is good in general" ($N=11$). Additionally, three participants commented the "natural headpose" ($N=3$) in the \textit{low\_static} condition. In contrast, the conditions with high base height received mostly negative comments like "too high" (static: $N=23$; dynamic: $N=21$), "the position of POIs is difficult to estimate" (static: $N=9$; dynamic: $N=9$), and "uncomfortable head position" (static: $N=6$; dynamic: $N=6$). 

For \textit{visibility}, no significant interactions were found, contradicting hypothesis $H_{H2}$ (A base height of approximately eye level leads to higher visibility). Interestingly, all settings were perceived to be equally visible, albeit lower base heights were generally preferred. There were limited comments about visibility and the comments that did touch on visibility presented contrasting viewpoints among participants. For instance, four participants expressed positive views, such as "my view of the POIs on the car's roof is unobstructed." Conversely, three participants conveyed a similar comment, only with a negative connotation with "the POIs are on the car's roof." 

The dynamic scaling of POIs did not have any significant effect on our participants ratings. While the dynamic scaling of POIs received positive feedback for circumventing occlusion-related issues at a distance, the chosen maximum height at a distance was deemed excessively high. This is reflected in comments such as "uncomfortable head position" ($N=5$). Thus, we reject hypothesis $H_{H3}$ (Dynamic height scaling leads to higher pleasantness).

In summary, a low base height approximately set at eye level is preferred by users, resulting in significantly higher satisfaction and comfort. Additionally, we recommend configuring dynamic height scaling based on user preferences. While dynamic scaling can potentially mitigate occlusion issues at a distance, it may come at the cost of an uncomfortable head position.

\subsection{Size}
Subsequently, we evaluated various POI scaling options, again considering two levels for base scale with or without dynamic scaling over distance. 

Based from our results for size in Section \ref{sec:sizeResults} confirm hypothesis $H_{S1}$ (A POI base size of approximately three meters leads to higher \textit{pleasantness}). However, the pleasantness ratings were still not optimal as the mean ratings for pleasantness did not exceed one of a maximum of three. This can be explained by the comments made by participants, as there was no clear consensus for a favorite size. Positive comments noted that the size in the \textit{small\_dynamic} condition was deemed "exactly right, fitting" ($N=10$). Conversely, some participants remarked on personal preference, citing POIs being "too small when far away" ($N=5$), "too big when far away" ($N=2$), "too big" ($N=3$), or "too small" ($N=2$). In the \textit{small\_static} condition, POIs were predominantly perceived as "too small" ($N=9$) or "too small when far away" ($N=9$). Consequently, text on POIs was deemed "difficult to read" ($N=5$) by certain participants. Nevertheless, other found this condition pleasing, describing it as "pleasant, exactly right" ($N=6$).

We reject $H_{S2}$ (POIs with a large base size of approximately 7.5 meters lead to higher \textit{visibility}). The large base size proved excessive for many participants, with comments for \textit{large\_static} including remarks such as "too big" ($N=11$), "too big when nearby" ($N=5$), and "POIs occlude the real world too much" ($N=4$). However, five participants favored this condition, and another three achknowledged the ease of reading text.
The \textit{large\_dynamic} exacerbated issues, with 19 participants indicating that the POIs were "too big" in general ($N=12$) or "too big when nearby" ($N=7$).

We confirm $H_{S3}$ (Dynamic size scaling leads to higher \textit{visibility} and \textit{pleasantness}). This could stem from better readability, as ten participants called the \textit{small\_dynamic} as "exactly right", and five participants calling the \textit{large\_dynamic} "exactly right". Dynamically sized POIs also don't seem to influence the participants depth perception, which is in line with the findings of Dey et al. \cite{dey2012tablet}, where correct depth estimation in handheld AR was mostly dependant on object height and not size.

In summary, we recommend to scale POIs at a radius of roughly three meters with distance-based dynamic scaling as proposed in Equation \ref{eq:scalingFormula} for good satisfaction, visibility, and pleasantness. In addition, some personalization should be considered. Each condition regulating the POIs' size got at least five positive remarks, as such making all conditions suitable for at least a subset of users. Some seem to favor less visual clutter of smaller POIs while some participants seem to favor bigger POIs for better readability.

\subsection{Rotation}
For rating the rotation of POIs we used word pairs for the categories clarity, support, complexity, and pleasantness. The billdboarding condition was rated significantly higher within each word pair. Thus we confirm hypothesis $H_{R1}$ (Billdboarding POIs are more supportive for delivering
information and are thus more pleasant), as they were rated significantly better in support and pleasantness. This was also mirrored in the participants comments during the study. Comment categories for \textit{billboarding} included "good" ($N=10$), "pleasant" ($N=8$), and "everythign is readable" ($N=4$). The only negatively coded comments for this condition were from three people, that said "the rotation is a little too late".

The \textit{no rotation} condition was rated significantly worse than \textit{billdboarding} in each category and received mostly neutral to negative ratings. Thus, we reject hypothesis $H_{R2}$ (Non-rotating POIs are easier to understand and give a clearer overview of the environment). Participants either did not like the three dimensional POIs or did not understand what information they could gain by them. As such, the condition \textit{no rotation} got mostly negative comments in lines of "i don't like that the POIs are visualized as spheres" ($N=9$), "useless" ($N=7$), "confusing" ($N=7$), "there is nothing gained for me when I see the spheres from behind" ($N=5$), and "unpleasant" ($N=4$). In addition, eleven participants directly told us that they liked the billboarding condition better. Nevertheless, six participants liked the "visualization as spheres". The comments also show, that some participants did perceive the billboarding spheres as 2D objects while the non-rotating spheres where perceived as 3D objects.

In summary, we clearly recommend for POIs to show billboarding rotation behavior, meaning they should always face the user. This leads, according to our data, to better clarity, support, complexity, and pleasantness compared to non-rotating POIs. Still, the obtained ratings were not optimal and other methods of rotation should be evaluated further. This could for example encompass rotating the POIs with slight anticipation for the cars heading vector. Another potential direction could be a small idling rotation in combination with a subtle hovering animation to give more of a 3D impression to the POIs without confusing users like with our static spheres.

\subsection{Render Distance}
While there was a significant effect of the render distance on the dependent variable ($t(37)=7.29; p<.001$), both conditions were suboptimal. The \textit{long distance} was rated to render POIs too early ($M=-0.789; Mdn=-0.5; SD=1.32$) while for the \textit{short distance}, POIs rendered too late ($M=-0.684; Mdn=1.0; SD=1.07$). As such, the optimal point to fade in POIs seems to be somewhere between 150 and 450 meters. Thus, we reject our hypothesis $H_{RD1}$: (A short render distance is preferred by users). 

This is also mirrored in the comments participants made. For the \textit{long distance}, participants liked the "good overview" ($N=8$) and the "information about restaurants farther away" ($N=4$). Though, nine participants commented about "information overload" and "too many POIs" ($N=8$). Additionally, six people complained, that POIs in the distance are not readable and that some POIs occluded each other ($N=2$). 

For the \textit{short distance}, participants positively noted the "lower quantity of POIs" ($N=8$), "pleasantness" ($N=6$), and "no information overload" ($N=5$). However, eight participants reported that the POIs get displayed "a little too late" and that there were "not enough POIs or limited options" ($N=5$). For both conditions there were four participants that emphasized that the render distance should be dependent on location and circumstances.

In summary, both the 150 meters and the 450 meters rendering distance were suboptimal for our conditions, namely, an industrial area with speed capped at 30 km/h. As such, a value in between our chosen values, e.g. 300 meters, could be fitting for the conditions employed in our study. However, this value has to adapted to the circumstances as well as to the users preferences.

\subsection{Information Density}
For the amount and types of content displayed on the face of POIs, participants clearly preferred our condition with more content with a significant main effect between the two conditions. The \textit{high information density} condition was close to ideal ($M=-0.378; Mdn=0; SD=0.861$) but still showed a trend of not enough information. Those results are also mirrored in the recorded comments, as the participants expressed mostly negative comments regarding the \textit{low information density}. These included "too little information" ($N=14$), "star rating is missing" ($N=9$), "image is missing" ($N=8$), and "ony the name is not enough" ($N=4$). For the \textit{high information density} condition, more positive comments were expressed, like "I like it" ($N=7$) and "informative" ($N=3$). Still, users wished for even more information displayed ($N=12$) or that the image and name only gets displayed for nearby POIs ($N=6$). For both conditions, four participants respectively mentioned the cut-off name of some restaurants. Thus, we partially confirm hypothesis $H_{I1}$ (POIs with three types of data are preferred by users), as our condition with three types of data received overall positive ratings. However, more information was wanted by some participants.

For the name, star rating and image, a trend can be identified for showing more information on nearby POIs and less information for far POIs, confirming hypothesis $H_{I2}$ (Users want to have more data displayed for nearby POIs). However, when to display the icon was polarizing, with 40\% requesting to always show the icon and with 40\% requesting to never show the icon. This could stem from the participants wishes to display the category of POIs, such as restaurants. As such, the icon visualizing the type of POI could be added to our existing three data types. On the contrary, 40\% never want the icon to displayed. From the user comments we can deduce, that some users rather want to have the image displayed at all times.

In the post-questionnaires, participants could also mention additional types of information that they would want to see on POIs. Those informations like "distance to the POI" ($N=7$) could be also displayed directly on the POIs. Another possibility would be to include a functionality to select POIs as proposed in \cite{Schramm23Assessing}. After selection, some of the user-commented categories could be shown, such as "business hours" ($N=11$), "reservations" ($N=5$), "price range" ($N=4$) or a "call button" ($N=4$).

In summary, we advice to at least display the POI name, rating and image always or when nearby. Furthermore, some other information such as the POI category should be displayed as well. In addition, POIs could be made selectable to then display more information like opening times.

\subsection{Appearance}
The first part of our post-questionnaires was focussed on the POIs appearance. Here, users could give additional information via text when they rated an aspect with a three or lower. Overall, the POIs general appearance and form were rated highly. Thus, we confirm hypothesis $H_{A1}$ (The visual presentation of our POIs is appealing). Comments explaining their negative rating regarding the appearance included: "I wish for more focus on the content" ($N=3$), "the border should be more subtle" ($N=3$), and "Some POIs are difficult to differentiate" ($N=2$). As for the color of POIs, the results were more polarizing with 37\% of participants rating the color three and lower, while 47\% rated the color a five and better. In negative comments participants called the color "too prominent" ($N=9$) or wished for more variety in colors ($N=3$). Thus, we can confirm $H_{A2}$ (The POIs' color makes them stand out against the environment) as the color indeed lets the POIs stand out. However, for some participants the color stood out too much and was too prominent as described by their comments and reflected in the ratings. In addition, participants could comment any aspects of the POIs' appearance that they would want to change. Eight participants wished, that the name of restaurants would be readable. This propably stems from the problem that some of our fake-POI data had long restaurant names which were then cut off due to limited allocated space for displaying the name. Six participants wanted a different form like a pin needle, a drop with the tip facing down, or a rectangular form for more text space. Four participants wanted for the POIs to differentiate more, such as color coded restaurant categories.

In summary, our POI design has promising aspects as the appearance and form were generally liked. Thus, round or spherical POIs with a futuristic look are a promising baseline. However, in future work, more designs, colors, forms and layouts should be explored. A promising approach would be to allow some user-customization of POIs' appearance, form, color, and content, potentially connected to the cars digital design in some way. Furthermore, the use of design elements to differentiate between kinds or categories of POIs could help with exploration and orientation.

\subsection{Preferred Appearance and Positioning}
Based on our results and discussion, we can answer research question \textbf{RQ1} (What appearance and positioning of POIs do users prefer in an in-car AR context). We recommend the following baselines for visualizing in-car POIs based on the user preference in our study:
\begin{itemize}
    \item \textbf{Height:} A low base-height set around eye-level with dis\-tance\--based vertical scaling depending on the context and the user's preference.
    \item \textbf{Size:} A scale of roughly 3 meters with a small amount of distance-based scaling.
    \item \textbf{Rotation:} Billboarding POIs, meaning a continuous rotation around the x-axis and y-axis in order to always face the user's head position.
    \item \textbf{Render Distance:} Far away POIs should fade out after a certain distance depending on the car's environment and speed. For urban areas with a speed of 30km/h we recommend to fade-out POIs at a distance of around 300 meters. However, more research in this area is needed.
    \item \textbf{Information Density:} POIs should at least display the location's name, a representative image, and a rating if applicable. In addition, POIs should somehow indicate their category such as restaurant or gas station, e.g. through an icon. More information based on the POIs category is welcome and should be made available in some way, e.g. through further interaction.
    \item \textbf{Appearance:} We recommend to base the POIs' appearance on other designs that are used in the space where the function is deployed. Users should be able to customize the color of POIs and should potentially also be able to customize the form. Spherical POIs with a futuristic design were rated positively in our scenario.
\end{itemize}

\subsection{Acceptance}
The in-car AR-function was generally accepted by users. Each statement in our acceptance questionnaire was rated in median with a five or higher. Notably the statements for regular use of the AR-function, both as vehicle owner and for other passengers, were rated positively with a median score of six. In addition, most participants agreed with a median score of six, that the AR-function excites them, that it is useful, and that they expect added value from it. Thus, based on our data, we can answer our research question \textbf{RQ2} (Do users accept an AR system inside a moving vehicle?) with yes. Our participants accepted the AR-system inside the moving vehicle with genereally positive results.

As seen in the comments in table \ref{tab:AcceptanceComments}, participants particularly liked the information gain about their environment and the help for searching, as suggested in \cite{BergerGridStudyInCarPassenger2021}. Negative comments mostly regarded the visual design and the information overload, both of which should be studied further based on the insights in this paper.

Regarding the occurrence of MS, none of the participants reported experiencing any symptoms, and no study runs were aborted due to MS.

\begin{table}[h]
    \centering
    \caption{Coded comment categories regarding what the participants liked or disliked about the AR-function in general.}
    \label{tab:AcceptanceComments}
    \begin{tabular}{l|l|c}
    \toprule
    \textbf{Question} & \textbf{Comment Category}  & \textbf{N}  \\
    \midrule
    Liked     & Information gain            & 18 \\
             & Search help                  & 8  \\
             & Innovative                   & 4  \\
             & Visual design                & 4  \\
             & Less distracting             & 3  \\
             & Other                        & 6  \\

    \midrule
    Disliked & Visual design                & 8  \\
             & Information overload         & 8  \\
             & Digital occludes real world  & 6  \\
             & Not enough information       & 4  \\
             & No added value               & 2  \\
             & Other                        & 8  \\
    \bottomrule  
    \end{tabular}
\end{table}

\subsection{Intention of Use}
\label{sec:intentionOfUse}
The participants intention of use was rather pragmatic, as chargepoints, gas stations and supermarkets were amongst the highest rated categories for POIs. Among the presented infotainment use-cases, only the "landmarks in unknown cities" use-case was rated rather highly with a mean score of 5.9. Thus we confirm the user acceptance of this infotainment use-case, as Berger et al. showed in their work \cite{BergerRearSeatDoor21}. For the additional comments, participants also mentioned parking ($N = 6$) and hospitals ($N = 5$), again pragmatic use-cases that can support them in their daily life.

In summary, we advise to focus on POI categories that may help users in their daily life for a helpful assistance system. Though, the use of POIs to entertain passengers during trips in unknown cities is also a promising approach.

\subsection{Limitations}
Given that our study was conducted in a singular environment with a fixed speed limit, the applicability of some of our results may be limited to this specific context and may not translate to other settings such as traffic-dense urban areas or highways. Especially some of our variables like the render distance or size would likely require adaptation to different contexts. Furthermore, our POI appearance was confined to a single spherical design, which could restrict the generalizability of our findings to alternative design choices. Aditionally, the chosen scenario itself could have influenced the results.

Additionally, the Varjo XR-3 hardware has some limitations. Participants wore the HMD continuously for 30 to 60 minutes, potentially leading to neck strain due to the device's weight. This factor could have influenced our results, particularly considering our emphasis on hedonic qualities like satisfaction and pleasantness. For comparison, a study by Kim et al. \cite{kim2021Discomfort} found that while physical discomfort and simulator sickness symptoms emerged approximately twice as quickly when using a VR headset compared to a desktop monitor, all participants were able to complete the 60-minute task without significant discomfort. Future research could explore a comparison between VST and optical see-through devices for in-car applications.

Also, as our study was conducted from the back seat, the transferability of our results to the front seats may be limited. All POIs would be visible in all seats, since the digital objects occluded the real world view and POIs were visible through all windows. Nevertheless, the back seat positions allow the user to view POIs above them through the ceiling while front seat positions allow unobstructed view of POIs through the windshield. This could especially affect the height and size conditions.

None of the participants reported any MS symptoms. However, measuring MS with quantitative data, such as a standardized questionnaire like the Motion Sickness Assessment Questionnaire (MSAQ) \cite{gianaros2001MSAQ}, would have offered more objective and insightful results. Additionally, recording eye-tracking data to quantify participants' focus on POIs could have also provided further valuable insights.

Lastly the weather and lighting conditions were not uniform across all study session, since we did a field study over multiple days. This may have influenced participants' perceptions and ratings.

\section{Conclusion and Future Work}
\label{section:Conclusion}
In this paper we presented \textit{Blending the Worlds}, a system for displaying points of interest (POIs) in the environment of a moving car via a pass-through head-mounted display. We explored various parameters of in-car Augmented Reality (AR) based POI visualization, including positioning, scaling, render distance, information density, POI appearance, and the acceptance of such an AR-function. We conducted a comprehensive user study in a moving vehicle under real-life conditions with 38 participants, collecting and analysing both quantitative and qualitative data. 

Based on our data, we answer which appearance and positioning of POIs users prefer in an in-car AR context (\textbf{RQ1}). For \textbf{height}, we recommend a vertical position set around eye-level with distance-based height scaling depending on the context and the user's preference. For \textbf{size}, we recommend scaling POIs at around three meters, deploying distance-based scaling with our proposed equation. For \textbf{render distance}, we advise to fade out far away POIs after a certain distance depending on the car's speed and environment. Our recommendation for urban areas with a speed limit of 30km/h is a fade-out distance of around 300 meters. Our users preferred a high \textbf{information density}, where the location's name, image, and rating should be displayed at most times. POIs also should indicate their category, e.g. through utilizing a fitting icon. As for \textbf{appearance}, we evaluated a futuristic design with spherical POIs which was generally rated positively, except for polarizing ratings regarding the pink border color. We recommend to further investigate other designs and to make the POIs' form and color customizable by the user.

We also address the question of whether users accept an AR system inside a moving vehicle (\textbf{RQ2}). Our results indicate a general acceptance of using POIs in the context of in-car AR, with users describing the system as innovative, useful, and value-adding. We recommend prioritizing POI categories that provide practical assistance in daily life, such as charging stations, gas stations, and supermarkets. Furthermore, utilizing POIs to entertain passengers by showcasing landmarks during trips in unfamiliar cities offers another promising avenue.

Future works could explore customization of POIs or investigate means of interaction to display more information and enable direct interaction with the locations, e.g. by calling them or making a reservation. We plan to integrate the system into more contexts and environments like urban areas, highways or along scenic routes while using real-world data. Investigating systems that place POIs directly on real-world objects utilizing computer vision also seems promising. Our system could also be expanded by investigating depth cues and occlusion of POIs for in-car AR. Additionally, the system could be tested in different seating positions or in other transportation methods like trains. Overall, we recommend to further explore concepts that utilize in-car AR, especially targeted towards passengers for information and entertainment use-cases.


\bibliographystyle{ACM-Reference-Format}
\bibliography{bibliography}

\clearpage

\appendix
\section{Appendix}

\subsection{Questionnaires}
\label{sec:AppendixQuestionnaires}

Our questions regarding the participants' affinity for AR. Three questions (Q0, Q1, and Q2) stem from Janzik \cite{janzik2022studie} where we applied slight modifications, while the other three questions (Q3, Q4, and Q5) were specifically crafted for our study.
\begin{itemize}
    \item \textbf{ar-Q0}: I would find using AR entertaining.
    \item \textbf{ar-Q1}: All in all, AR is a good technology.
    \item \textbf{ar-Q2}: It would be easy for me to become experienced in using AR.
    \item \textbf{ar-Q3}: I am positive towards new technologies such as AR.
    \item \textbf{ar-Q4}: I think the use of AR in vehicles sould make sense.
    \item \textbf{ar-Q5}: I would like to use AR functions in vehicles.
\end{itemize}

The height and size conditions both used three questions with a seven point likert scale for each questions, ranging from -3 to +3.
\begin{itemize}
    \item \textbf{Satisfaction}: Here, -3 corresponded to \textit{excessively low or small}; 0 to \textit{exactly right}; and +3 to \textit{excessively high or big}
    \item \textbf{Strikingness}: Here, -3 corresponded to \textit{excessively covert}; 0 to \textit{exactly right}; and +3 to \textit{excessively overt}
    \item \textbf{Pleasantness}: Here, -3 corresponded to \textit{very unpleasant}; 0 to \textit{mixed}; and +3 to \textit{very pleasant}
\end{itemize}

For the rating of the rotation, we used four word pairs, again with a seven point likert scale ranging from -3 to +3. The word pairs for clarity, support, and complexity were taken from the short version of the User Experience Questionnaire (UEQ-S) \cite{schrepp2017design}:
\begin{itemize}
    \item \textbf{Clarity}: -3 = confusing; +3 = clear
    \item \textbf{Support}: -3 = obstructive; +3 = supportive
    \item \textbf{Complexity}: -3 = complicated; +3 = easy
    \item \textbf{Pleasentness}: -3 = unpleasant; +3 = pleasant
\end{itemize}

For render distance and information density, participants could only rate one scale for each condition:
\begin{itemize}
    \item \textbf{Visibility (render distance)}: where -3 meant excessively early, 0 meant exactly right, and +3 meant excessively late.
    \item \textbf{Information density}: where -3 meant way too little information, 0 meant exactly right, and +3 meant way too much information.
\end{itemize}

After the study procedure in the car, participants filled out the post-questionnaires. First, they could rate the AR function in general. Each of the questions utilized a seven point likert scale, where 1 corresponded to \textit{strongly disagree} and 7 corresponded to \textit{strongly agree}. The question order was randomized for each participant:
\begin{itemize}
    \item \textbf{postAr-Q0} The AR function increased my comfort.
    \item \textbf{postAr-Q1} The AR function is useful.
    \item \textbf{postAr-Q2} The AR function is innovative.
    \item \textbf{postAr-Q3} I would purchase the AR function as an optional feature.
    \item \textbf{postAr-Q4} I am confident that it will function reliably and without errors.
    \item \textbf{postAr-Q5} The AR function excites me.
    \item \textbf{postAr-Q6} The AR function satisfies my need to experience something new.
    \item \textbf{postAr-Q7} If I had this AR function in the vehicle, I would use it regularly.
    \item \textbf{postAr-Q8} My passengers would appreciate it if I had this in my car.
    \item \textbf{postAr-Q9} My passengers would use it regularly if I had this AR function in my car.
    \item \textbf{postAr-Q10} I expect added value from this AR function.
\end{itemize}

The next questions regarded the POIs appearance, form, and color with seven point likert scales ranging from 1: \textit{very dissatisfied} to 7: \textit{very satisfied}. Additionally, there were two freetext fields with the questions: 
\begin{itemize}
    \item \textbf{postPoi-Q0}: Are there any other aspects regarding the design of the POIs that you would like to change?
    \item \textbf{postPoi-Q1}: Are there any additional pieces of information you would like to add to the POIs?
\end{itemize}

\subsection{Height}
\begin{table}[ht]
    \centering
    \caption{The mean, median and standard deviation values for the height ratings, grouped by condition. For height and Visibility, zero is best. For Pleasantness, higher is better.}
    \begin{tabular}{llrrr}
    \toprule
    \textbf{Condition} & \textbf{Dep. Variable} & \textbf{Mean}            & \textbf{Median}       & \textbf{SD}              \\
    \midrule
    low\_static        & Satisfaction                & -0.395                   & 0                     & 1                        \\
                       & Visibility                  & 0.474                    & 0                     & 0.979                    \\
                       & Pleasantness                & 0.316                    & 0                     & 1.44                     \\
    low\_dynamic       & Satisfaction                & -0.211                   & 0                     & 1.14                     \\
                       & Visibility                  & 0.579                    & 0.5                   & 1.18                     \\
                       & Pleasantness                & 0.474                    & 0                     & 1.5                      \\
    high\_static       & Satisfaction                & 1.66                     & 2                     & 1.36                     \\
                       & Visibility                  & 0.289                    & 0                     & 1.14                     \\
                       & Pleasantness                & -0.447                   & 0                     & 1.37                     \\
    high\_dynamic      & Satisfaction                & 1.76                     & 2                     & 1.24                     \\
                       & Visibility                  & 0.211                    & 0                     & 1.02                     \\
                       & Pleasantness                & -0.289                   & 0                     & 1.47                     \\
    \bottomrule
    \end{tabular}
\end{table}

\begin{table*}[ht]
    \centering
    \caption{Results for the repeated measures ANOVA for the independent variable \textit{height} in the \textit{satisfaction} condition.}
    \begin{tabular}{r|cccccc}
    \toprule
                                  & \textbf{Sum of Squares} & \textbf{df} & \textbf{Mean Square} & \textbf{F} & \textbf{p} & \textbf{$\eta^2p$} \\
    \midrule
    Base Height                   & 154.0066                & 1           & 154.0066             & 98.6821    & .001       & 0.727              \\
    Residual                      & 57.7434                 & 37          & 1.5606               &            &            &                    \\
    Dynamicity                    & 0.7961                  & 1           & 0.7961               & 0.9515     & 0.336      & 0.025              \\
    Residual                      & 30.9539                 & 37          & 0.8366               &            &            &                    \\
    Base Height * Dynamicity      & 0.0592                  & 1           & 0.0592               & 0.0401     & 0.842      & 0.001              \\
    Residual                      & 54.6908                 & 37          & 1.4781               &            &            &                    \\
    \bottomrule
    \multicolumn{7}{l}{\textit{Note. Type 3 Sums of Squares}}                                                                                                                                                                                                  
    \end{tabular}
\end{table*}

\begin{table*}[ht]
    \centering
    \caption{\textit{Post hoc} Comparisons - \textit{Base Height} $\ast$ \textit{Dynamicity} for the independent variable \textbf{height} in the \textbf{satisfaction} condition. Siginifcant p-values are highlighted in bold.}
    \begin{tabular}{lllll|ccccc}
    \toprule
    \textbf{Base Height} & \textbf{Dynamicity} & -  & \textbf{Base Height} & \textbf{Dynamicity} & \textbf{Mean Difference} & \textbf{SE} & \textbf{df} & \textbf{t} & \textbf{ptukey} \\
    \midrule
    Low                  & Static              & - & Low                   & Dynamic             & -0.184                   & 0.232       & 37.0        & -0.794     & 0.857           \\
                         &                     & - & High                  & Static              & -2.053                   & 0.287       & 37.0        & -7.149     & \textbf{<.001}                      \\
                         &                     & - & High                  & Dynamic             & -2.158                   & 0.263       & 37.0        & -8.213     & \textbf{<.001}                      \\
                         & Dynamic             & - & High                  & Static              & -1.868                   & 0.239       & 37.0        & -7.816     & \textbf{<.001}                      \\
                         &                     & - & High                  & Dynamic             & -1.974                   & 0.278       & 37.0        & -7.089     & \textbf{<.001}                      \\
    High                 & Static              & - & High                  & Dynamic             & -0.105                   & 0.261       & 37.0        & -0.404     & 0.977             \\
    \bottomrule                           
    \end{tabular}
\end{table*}

\begin{table*}[ht]
    \centering
    \caption{Results for the repeated measures ANOVA for the independent variable \textit{height} in the \textit{visibility} condition.}
    \begin{tabular}{r|cccccc}
    \toprule
                                  & \textbf{Sum of Squares} & \textbf{df} & \textbf{Mean Square} & \textbf{F} & \textbf{p} & \textbf{$\eta^2p$} \\
    \midrule
    Base Height                   & 2.90132                 & 1           & 2.90132              & 3.3185     & 0.077      & 0.082              \\
    Residual                      & 32.34868                & 37          & 0.87429              &            &            &                    \\
    Dynamicity                    & 0.00658                 & 1           & 0.00658              & 0.0150     & 0.903      & 0.000              \\
    Residual                      & 16.24342                & 37          & 0.43901              &            &            &                    \\
    Base Height * Dynamicity      & 0.32237                 & 1           & 0.32237              & 0.5699     & 0.455      & 0.015              \\
    Residual                      & 20.92763                & 37          & 0.56561              &            &            &                    \\
    \bottomrule
    \multicolumn{7}{l}{\textit{Note. Type 3 Sums of Squares}}                                                                                                                                                                                                  
    \end{tabular}
\end{table*}

\begin{table*}[ht]
    \centering
    \caption{\textit{Post hoc} Comparisons - \textit{Base Height} $\ast$ \textit{Dynamicity} for the independent variable \textbf{height} in the \textbf{visibility} condition. Siginifcant p-values are highlighted in bold.}
    \begin{tabular}{lllll|ccccc}
    \toprule
    \textbf{Base Height} & \textbf{Dynamicity} & -  & \textbf{Base Height} & \textbf{Dynamicity} & \textbf{Mean Difference} & \textbf{SE} & \textbf{df} & \textbf{t} & \textbf{ptukey} \\
    \midrule
    Low            & Static           & - & Low            & Dynamic          & -1.053                   & -0.1053       & 37.0        & -0.598     & 0.932           \\
                   &                  & - & High            & Static           & 0.1842                   & 0.192       & 37.0        & 0.961     & 0.772                      \\
                   &                  & - & High            & Dynamic          & 0.2632                   & 0.163       & 37.0        & 1.614     & 0.383                      \\
                   & Dynamic          & - & High            & Static           & 0.2895                   & 0.206       & 37.0        & 1.403     & 0.505                      \\
                   &                  & - & High            & Dynamic          & 0.3684                   & 0.197       & 37.0        & 1.865     & 0.260                      \\
    High            & Static           & - & High            & Dynamic          & 0.0789                   & 0.148       & 37.0        & 0.534     & 0.950             \\
    \bottomrule                           
    \end{tabular}
\end{table*}

\begin{table*}[ht]
    \centering
    \caption{Results for the repeated measures ANOVA for the independent variable \textit{height} in the \textit{visibility} condition.}
    \begin{tabular}{r|cccccc}
    \toprule
                                  & \textbf{Sum of Squares} & \textbf{df} & \textbf{Mean Square} & \textbf{F} & \textbf{p} & \textbf{$\eta^2p$} \\
    \midrule
    Base Height                   & 22.132                  & 1           & 22.132              & 7.591       & 0.009      & 0.170              \\
    Residual                      & 107.868                 & 37          & 2.915               &             &            &                    \\
    Dynamicity                    & 0.947                   & 1           & 0.947               & 0.854       & 0.361      & 0.023              \\
    Residual                      & 41.053                  & 37          & 1.110               &             &            &                    \\
    Base Height * Dynamicity      & 0.000                   & 1           & 0.000               & 0.000       & 1.000      & 0.000              \\
    Residual                      & 46.000                  & 37          & 1.243               &             &            &                    \\
    \bottomrule
    \multicolumn{7}{l}{\textit{Note. Type 3 Sums of Squares}}                                                                                                                                                                                                  
    \end{tabular}
\end{table*}

\begin{table*}[ht]
    \centering
    \caption{\textit{Post hoc} Comparisons - \textit{Base Height} $\ast$ \textit{Dynamicity} for the independent variable \textbf{height} in the \textbf{pleasantness} condition. Siginifcant p-values are highlighted in bold.}
    \begin{tabular}{lllll|ccccc}
    \toprule
    \textbf{Base Height} & \textbf{Dynamicity} & -  & \textbf{Base Height} & \textbf{Dynamicity} & \textbf{Mean Difference} & \textbf{SE} & \textbf{df} & \textbf{t} & \textbf{ptukey} \\
    \midrule
    Low                  & Static              & - & Low                   & Dynamic             & -0.158                   & 0.254       & 37.0        & -0.620     & 0.925           \\
                         &                     & - & High                  & Static              & 0.763                    & 0.288       & 37.0        & 2.647      & 0.055           \\
                         &                     & - & High                  & Dynamic             & 0.605                    & 0.326       & 37.0        & 1.859      & 0.263           \\
                         & Dynamic             & - & High                  & Static              & 0.921                    & 0.325       & 37.0        & 2.832      & \textbf{0.036}  \\
                         &                     & - & High                  & Dynamic             & 0.763                    & 0.368       & 37.0        & 2.071      & 0.181           \\
    High                 & Static              & - & High                  & Dynamic             & -0.158                   & -0.158      & 37.0        & -0.650     & 0.915           \\
    \bottomrule                           
    \end{tabular}
\end{table*}

\clearpage
\onecolumn

\subsection{Size}
\begin{table*}[ht]
    \centering
    \caption{The mean, median and standard deviation values for the size ratings, grouped by condition. For size and visibility, zero is best. For Pleasantness, higher is better.}
    \begin{tabular}{llrrr}
    \toprule
    \textbf{Condition} & \textbf{Dependent Variable} & \textbf{Mean} & \textbf{Median} & \textbf{SD} \\
    \midrule
    small\_static      & Satisfaction           & 1.74          & 2               & 1.267       \\
                       & Visibility             & 1.63          & 2               & 1.217       \\
                       & Pleasantness           & -1.00         & -1              & 1.115       \\
    small\_dynamic     & Satisfaction           & -0.05         & 0               & 1.012       \\
                       & Visibility             & 1.53          & 2               & 1.350       \\
                       & Pleasantness           & 1.39          & 1               & 1.306       \\
    large\_static      & Satisfaction           & -0.50         & 0               & 0.893       \\
                       & Visibility             & 0.21          & 0               & 0.991       \\
                       & Pleasantness           & -0.45         & -1              & 1.465       \\
    large\_dynamic     & Satisfaction           & -0.29         & -1              & 1.374       \\
                       & Visibility             & 0.21          & 0               & 1.630       \\
                       & Pleasantness           & 0.53          & 0.5             & 1.447       \\
    \bottomrule
    \end{tabular}
\end{table*}
\clearpage

\begin{table*}[ht]
    \centering
    \caption{Results for the repeated measures ANOVA for the independent variable \textit{size} in the \textit{satisfaction} condition.}
    \begin{tabular}{r|cccccc}
    \toprule
                                  & \textbf{Sum of Squares} & \textbf{df} & \textbf{Mean Square} & \textbf{F} & \textbf{p} & \textbf{$\eta^2p$} \\
    \midrule
    Base Size                     & 185.68                  & 1           & 185.684             & 143.7       & \textbf{<.001}      & 0.795              \\
    Residual                      & 47.82                   & 37          & 1.292               &             &            &                    \\
    Dynamicity                    & 6.74                    & 1           & 6.737               & 12.6        & \textbf{0.001}      & 0.254              \\
    Residual                      & 19.76                   & 37          & 0.534               &             &            &                    \\
    Base Size * Dynamicity        & 10.53                   & 1           & 10.526              & 19.5        & \textbf{<.001}      & 0.345              \\
    Residual                      & 19.97                   & 37          & 0.540               &             &            &                    \\
    \bottomrule
    \multicolumn{7}{l}{\textit{Note. Type 3 Sums of Squares}}                                                                                                                                                                                                  
    \end{tabular}
\end{table*}

\begin{table*}[ht]
    \centering
    \caption{\textit{Post hoc} Comparisons - \textit{Base Size} $\ast$ \textit{Dynamicity} for the independent variable \textbf{size} in the \textbf{satisfaction} condition. Siginifcant p-values are highlighted in bold.}
    \begin{tabular}{lllll|ccccc}
    \toprule
    \textbf{Base Height} & \textbf{Dynamicity} & -  & \textbf{Base Height} & \textbf{Dynamicity} & \textbf{Mean Difference} & \textbf{SE} & \textbf{df} & \textbf{t} & \textbf{ptukey} \\
    \midrule
    Small                & Static              & - & Small                  & Dynamic             & -0.947                   & 0.160       & 37.0        & -5.929     & \textbf{<.001}           \\
                         &                     & - & Large                  & Static              & -2.737                   & 0.225       & 37.0        & -12.148    & \textbf{<.001}           \\
                         &                     & - & Large                  & Dynamic             & -2.632                   & 0.237       & 37.0        & -11.113    & \textbf{<.001}           \\
                         & Dynamic             & - & Large                  & Static              & -1.789                   & 0.200       & 37.0        & -8.941     & \textbf{<.001}           \\
                         &                     & - & Large                  & Dynamic             & -1.684                   & 0.214       & 37.0        & -7.881     & \textbf{<.001}           \\
    Large                & Static              & - & Large                  & Dynamic             & 0.105                    & 0.176       & 37.0        & 0.598      & 0.932           \\
    \bottomrule                           
    \end{tabular}
\end{table*}

\begin{table*}[ht]
    \centering
    \caption{Results for the repeated measures ANOVA for the independent variable \textit{size} in the \textit{visibility} condition.}
    \begin{tabular}{r|cccccc}
    \toprule
                                  & \textbf{Sum of Squares} & \textbf{df} & \textbf{Mean Square} & \textbf{F} & \textbf{p} & \textbf{$\eta^2p$} \\
    \midrule
    Base Size                     & 97.92                   & 1           & 97.921              & 85.09       & \textbf{<.001}      & 0.697              \\
    Residual                      & 42.58                   & 37          & 1.151               &             &            &                    \\
    Dynamicity                    & 3.18                    & 1           & 3.184               & 5.80        & \textbf{0.021}      & 0.135              \\
    Residual                      & 20.32                   & 37          & 0.549               &             &            &                    \\
    Base Size * Dynamicity        & 6.74                    & 1           & 6.737               & 8.10        & \textbf{0.009}      & 0.180              \\
    Residual                      & 30.76                   & 37          & 0.831               &             &            &                    \\
    \bottomrule
    \multicolumn{7}{l}{\textit{Note. Type 3 Sums of Squares}}                                                                                                                                                                                                  
    \end{tabular}
\end{table*}

\begin{table*}[ht]
    \centering
    \caption{\textit{Post hoc} Comparisons - \textit{Base Size} $\ast$ \textit{Dynamicity} for the independent variable \textbf{size} in the \textbf{visibility} condition. Siginifcant p-values are highlighted in bold.}
    \begin{tabular}{lllll|ccccc}
    \toprule
    \textbf{Base Size}   & \textbf{Dynamicity} & -  & \textbf{Base Size}    & \textbf{Dynamicity} & \textbf{Mean Difference} & \textbf{SE} & \textbf{df} & \textbf{t} & \textbf{ptukey} \\
    \midrule
    Small                & Static              & - & Small                  & Dynamic             & -0.711                   & 0.203       & 37.0        & -3.504     & \textbf{0.006}           \\
                         &                     & - & Large                  & Static              & -2.026                   & 0.234       & 37.0        & -8.660     & \textbf{<.001}           \\
                         &                     & - & Large                  & Dynamic             & -1.895                   & 0.219       & 37.0        & -8.642     & \textbf{<.001}           \\
                         & Dynamic             & - & Large                  & Static              & -1.316                   & 0.203       & 37.0        & -6.467     & \textbf{<.001}           \\
                         &                     & - & Large                  & Dynamic             & -1.184                   & 0.223       & 37.0        & -5.318     & \textbf{<.001}           \\
    Large                & Static              & - & Large                  & Dynamic             & 0.132                    & 0.178       & 37.0        & 0.741      & 0.880           \\
    \bottomrule                           
    \end{tabular}
\end{table*}

\begin{table*}[ht]
    \centering
    \caption{Results for the repeated measures ANOVA for the independent variable \textit{size} in the \textit{pleasantness} condition.}
    \begin{tabular}{r|cccccc}
    \toprule
                                  & \textbf{Sum of Squares} & \textbf{df} & \textbf{Mean Square} & \textbf{F} & \textbf{p} & \textbf{$\eta^2p$} \\
    \midrule
    Base Size                     & 20.632                  & 1           & 20.632              & 4.898       & \textbf{0.033}      & 0.117              \\
    Residual                      & 155.868                 & 37          & 4.213               &             &                     &                    \\
    Dynamicity                    & 2.132                   & 1           & 2.132               & 2.685       & 0.110               & 0.068              \\
    Residual                      & 29.368                  & 37          & 0.794               &             &                     &                    \\
    Base Size * Dynamicity        & 0.237                   & 1           & 0.237               & 0.189       & 0.666               & 0.005              \\
    Residual                      & 46.263                  & 37          & 1.250               &             &                     &                    \\
    \bottomrule
    \multicolumn{7}{l}{\textit{Note. Type 3 Sums of Squares}}                                                                                                                                                                                                  
    \end{tabular}
\end{table*}

\begin{table*}[ht]
    \centering
    \caption{\textit{Post hoc} Comparisons - \textit{Base Size} $\ast$ \textit{Dynamicity} for the independent variable \textbf{size} in the \textbf{pleasantness} condition. Siginifcant p-values are highlighted in bold.}
    \begin{tabular}{lllll|ccccc}
    \toprule
    \textbf{Base Size}   & \textbf{Dynamicity} & -  & \textbf{Base Size}    & \textbf{Dynamicity} & \textbf{Mean Difference} & \textbf{SE} & \textbf{df} & \textbf{t} & \textbf{ptukey} \\
    \midrule
    Small                & Static              & - & Small                  & Dynamic             & -0.316                   & 0.253       & 37.0        & -1.247    & 0.602            \\
                         &                     & - & Large                  & Static              & 0.658                    & 0.392       & 37.0        & 1.676     & 0.350            \\
                         &                     & - & Large                  & Dynamic             & 0.500                    & 0.384       & 37.0        & 1.302     & 0.568            \\
                         & Dynamic             & - & Large                  & Static              & 0.974                    & 0.340       & 37.0        & 2.860     & 0.034            \\
                         &                     & - & Large                  & Dynamic             & 0.816                    & 0.365       & 37.0        & 2.233     & 0.133            \\
    Large                & Static              & - & Large                  & Dynamic             & -0.158                   & 0.208       & 37.0        & -0.758    & 0.873            \\
    \bottomrule                           
    \end{tabular}
\end{table*}

\twocolumn
\clearpage
\subsection{Rotation}
\begin{table}[H]
    \centering
    \caption{The mean, median and standard deviation values for the rotation word pair ratings, grouped by condition. Higher is better.}
    \begin{tabular}{llrrr}
    \toprule
    \textbf{Condition} & \textbf{Word Pair} & \textbf{Mean} & \textbf{Median} & \textbf{SD} \\
    \midrule
    Billboarding       & Clarity            & 1.55          & 2.00            & 1.25 \\
                       & Support            & 1.34          & 1.00            & 1.17 \\
                       & Complexity         & 1.55          & 2.00            & 1.16 \\
                       & Pleasantness       & 1.32          & 2.00            & 1.32 \\
    \midrule      
    No Rotation        & Clarity            & -0.474        & -1.00           & 1.75 \\
                       & Support            & -0.632        & -1.00           & 1.63 \\
                       & Complexity         & 0.0789        & -1.00           & 1.68 \\
                       & Pleasantness       & -0.868        & -1.00           & 1.68 \\
    \bottomrule
    \end{tabular}
\end{table}

\begin{table}[H]
    \centering
    \caption{T-test results for both rotation conditions.}
    \label{tab:ttestRotation}
    \begin{tabular}{llrr}
    \toprule
    \textbf{Category} & \textbf{T} & \textbf{df} & \textbf{p}       \\
    \midrule
    Clarity           & -6.908     & 37          & \textbf{$<.001$} \\
    Support           & -6.233     & 37          & \textbf{$<.001$} \\
    Complexity        & -4.581     & 37          & \textbf{$<.001$} \\
    Pleasantness      & -7.083     & 37          & \textbf{$<.001$} \\
    \bottomrule     
    \end{tabular}
\end{table}

\subsection{Render Distance}
\begin{table}[H]
    \centering
    \caption{The mean, median and standard deviation values for the render distance rating by condition. Closer to zero is better.}
    \begin{tabular}{lrrr}
    \toprule
    \textbf{Condition} & \textbf{Mean} & \textbf{Median} & \textbf{SD} \\
    \midrule
    Long Render Range  & -0.789        & -0.500          & 1.32        \\
    Short Render Range & 0.684         & 1.00            & 1.07        \\
    \bottomrule
    \end{tabular}
\end{table}

\subsection{Information Density}
\begin{table}[H]
    \centering
    \caption{The mean, median and standard deviation values for the information density ratings by condition. Closer to zero is better.}
    \begin{tabular}{lrrr}
    \toprule
    \textbf{Condition} & \textbf{Mean} & \textbf{Median} & \textbf{SD} \\
    \midrule
    Low Information Density  & -2.27   & -3.00           & 0.990        \\
    High Information Density & -0.378  & 0.00            & 0.861        \\
    \bottomrule
    \end{tabular}
\end{table}

\subsection{Appearance}
\begin{table}[H]
    \centering
    \caption{POI Appearance, 1-7, higher is better.}
    \label{tab:POIAppearance}
    \begin{tabular}{lccc}
    \toprule
    \textbf{Condition}   & Mean           & Mdn           & SD      \\
    \midrule
    \textbf{Appearance}  & 5.11           & 5.00          & 1.23    \\
    \textbf{Form}        & 5.21           & 6.00          & 1.19    \\
    \textbf{Color}       & 4.32           & 4.00          & 1.58    \\
    \bottomrule                
    \end{tabular}
\end{table}

\begin{table}[H]
    \centering
    \caption{Results for the post-questionnaire for AR appearance. Higher is better.}
    \label{tab:PostArGeneral}
    \begin{tabular}{lccc}
    \toprule
    \textbf{Question}                   & \textbf{Mean}  & \textbf{Mdn}   & \textbf{SD}   \\
    \midrule
    postAr-Q0 (Comfort)                 & 5.55           & 5.5            & 1.179         \\
    postAr-Q1 (Useful)                  & 5.68           & 6.0            & 1.210         \\
    postAr-Q2 (Innovative)              & 6.34           & 6.0            & 0.815         \\
    postAr-Q3 (Optional equipment)      & 4.84           & 5.0            & 1.701         \\
    postAr-Q4 (Doubt-free)              & 5.24           & 5.5            & 1.250         \\
    postAr-Q5 (Enthusiastic)            & 5.71           & 6.0            & 1.293         \\
    postAr-Q6 (Experiencing new things) & 5.37           & 5.0            & 1.172         \\
    postAr-Q7 (Regular use)             & 5.29           & 6.0            & 1.523         \\
    postAr-Q8 (Passengers)              & 5.76           & 6.0            & 1.324         \\
    postAr-Q9 (Utilizing passengers)    & 5.58           & 6.0            & 1.426         \\
    postAr-Q10 (Added value)             & 5.84           & 6.0            & 1.305         \\
    \bottomrule    
    \end{tabular}
\end{table}

\end{document}